\documentclass[12pt,preprint]{aastex}
\usepackage{natbib}
\usepackage{graphicx}
\usepackage{amssymb}

\newcommand{\smallhi}{\mbox{H\,\tiny I}}
\newcommand{\hi}{\mbox{H\,\small I}}
\newcommand{\kms}{\mbox{km\ s}$^{-1}$}
\newcommand{\W}{W_{50}^{\rm obs}}
\newcommand{\vhel}{v_{\rm{hel}}}
\newcommand{\subhi}{_{\rm{\scriptstyle{H}\,\scriptscriptstyle I}}}
\newcommand{\Mhi}{M\subhi}
\newcommand{\Fhi}{F\subhi}

\newcommand{\df}{\mbox{DEF}\subhi}
\newcommand{\alf}{\mbox{ALFALFA}}
\newcommand{\msun}{M_\odot}
\newcommand{\ici}{C_{59}}
\newcommand{\rici}{C_{59,r}}

\slugcomment{Accepted for publication in The Astrophysical Journal}
\shorttitle{\hi\ CONTENT AND OPTICAL PROPERTIES OF \alf\ GALAXIES. I.}
\shortauthors{TORIBIO ET AL.}

\begin{document}

\title{\hi\ CONTENT AND OPTICAL PROPERTIES OF FIELD GALAXIES FROM THE \alf\
SURVEY. I.\ SELECTION OF A CONTROL SAMPLE}

\author{M.\ Carmen Toribio and Jos\'e M.\ Solanes}
\affil{Departament d'Astronomia i Meteorologia and Institut de Ci\`encies del
Cosmos, Universitat de Barcelona, Mart\'{\i} i Franqu\`es 1,
08028~Barcelona, Spain}
\email{mctoribio@am.ub.es, jm.solanes@ub.edu}
\and
\author{Riccardo Giovanelli and Martha P.\ Haynes}
\affil{Center for Radiophysics and Space Research, Space Sciences Building,
Cornell University, Ithaca, NY 14853}
\affil{National Astronomy and  Ionosphere Center,
 Cornell University, Ithaca, NY 14853 }
\email{riccardo@astro.cornell.edu, haynes@astro.cornell.edu}
\and
\author{Karen L.\ Masters} 
\affil{Institute of Cosmology and Gravitation, University of Portsmouth, Dennis Sciama Building, Burnaby Road, Portsmouth, P01 3FX, UK}
\email{karen.masters@port.ac.uk}

\begin{abstract}

  We report results from a study of the \hi\ content and stellar
  properties of nearby galaxies detected by the Arecibo Legacy Fast
  ALFA blind 21-cm line survey and the Sloan Digital Sky Survey. We
  consider two declination strips covering a total area of 9 hr
  $\times$ 16 deg in the general direction of the Virgo Cluster. The
  present analysis focuses on gas-rich galaxies expected to show
  little or no evidence of interaction with their surroundings. We
  seek to assemble a control sample suitable for providing absolute
  measures of the \hi\ content of gaseous objects, as well as to study
  the relationship between \hi\ emission and widely-used optical
  measures of morphology. From a database which includes more than
  $15,000$ \hi\ detections, we have assembled three samples that could
  provide adequate \hi\ standards. The most reliable results are
  obtained with a sample of 5647 sources found in low density
  environments, as defined by a nearest neighbor approach. The other
  two samples contain several hundred relatively isolated galaxies
  each, as determined from standard isolation algorithms based either
  on a combination of spectroscopic and photometric information or
  solely on photometric data. We find that isolated objects are not
  particularly gas-rich compared to their low-density-environment
  counterparts, while they suffer from selection bias and span a
  smaller dynamic range. All this makes them less suitable for
  defining a reference for \hi\ content. We have explored the optical
  morphology of gaseous galaxies in quiet environments finding that,
  within the volume surveyed, the vast majority of them display
  unequivocal late-type galaxy features. In contrast, bona fide
  gas-rich early-type systems account only for a negligible fraction
  of the 21-cm detections. We argue that \hi\ emission provides the
  most reliable way to determine the morphological population to which
  a galaxy belongs. We have also observed that the color distribution
  of flux-limited samples of optically-selected field \hi\ emitters
  does not vary significantly with increasing distance, while that of
  non-detections becomes notably redder. This result suggests that the
  colors and \hi\ masses of gas-rich galaxies cannot be very closely
  related.

\end{abstract}

\maketitle

\section{INTRODUCTION}

One of the defining characteristics of recent extragalactic research
is the execution of wide-area galaxy surveys, which have in many ways
revolutionized our understanding of both large-scale structure and
galaxy evolution. Most of them have been conducted in the
optical/near-infrared regime, where stellar light
dominates. Multicolor imaging surveys of virtually the entire sky,
some of which contain over a million extended sources, notably the Two
Micron All Sky Survey \citep[2MASS;][]{Skr06}, have been complemented
by spectroscopic surveys, such as the fifth edition of the Sloan
Digital Sky Survey (SDSS) Quasar Catalog \citep{Sch10}, based upon the
SDSS Seventh Data Release (DR7), which ranges in redshift from the
local universe to near the reionization epoch ($z\lesssim 6$) when the
first galaxies may have appeared.

The improvement in technological capabilities over the past decade has
enabled the astronomical community to build large homogeneous samples
of extragalactic objects in other regions of the electromagnetic
spectrum as well ---the radio wavelength regime being one of the most
prolific in this respect---, allowing us to develop a more thorough
account of the extragalactic realm. While early blind neutral hydrogen
(\hi) surveys, such as the Arecibo \hi\ Strip Survey \citep{Zwa97} or
the Arecibo Dual-Beam Survey \citep{RS02}, yielded useful estimates of
the \hi\ mass function from shallow samples containing a relatively
modest number of detections, the advent of focal plane arrays at 21-cm
enabled to carry out wide solid angle surveys that led to a
significant increase in the depth and, especially, size of the
databases. Between 1997 and 2001, the \hi\ Parkes All-Sky Survey
\citep[HIPASS;][]{Mey04} used a thirteen-beam array to sample the
entire southern sky and the northern up to $\mbox{Decl.} < + 25\degr$,
with a beam of 15.5 arcmin and a sensitivity of 17 mJy per beam at a
spectral resolution of 18 \kms. With a velocity range of $-1280$ to
$12,700$ \kms\ and more than 4000 \hi\ sources identified in the
southern sky alone, HIPASS was the first large untargeted \hi\ survey
that approached the dimensions necessary for a fair sampling of the
local volume.

Currently, a seven-beam array named ALFA (Arecibo L-band Feed Array),
installed at the Arecibo telescope in 2004, is being used to carry out
the Arecibo Legacy Fast ALFA Survey \citep[\alf;][]{Gio05}. Its aim is
to map, in the \hi\ line, 7000 square degrees of the high galactic
latitude sky between $0\degr$ and $+36\degr$ in declination. \alf\
offers significantly higher angular ($\mbox{FWHM} \sim 3\farcm 5$) and
spectral ($\sim 5.5$ \kms) resolution, as well as higher sensitivity
($\sigma_{\rm rms}=2.2$ mJy/beam at 10 \kms\ resolution) than any
previous survey of its kind. It is also deeper, with a radial extent
($-2000\lesssim cz_{\rm{hel}}\lesssim 18,000$ \kms) and a median depth
($cz_{\rm{hel}}\simeq 8000$ \kms) that should be sufficient to sample
satisfactorily the largest structures seen in numerical simulations in
a $\Lambda$CDM world model ($\sim 100\,h^{-1}$~Mpc). While HIPASS
detected 1 galaxy every $\sim 5$ square degrees, \alf\ is detecting
$\sim 5$ galaxies deg$^{-2}$. The current detection rate suggests that
\alf\ will have detected over 30,000 extragalactic \hi\ line
sources when completed. An account of the \alf\ survey goals and
technical characteristics, as well as its main contributions to science,
can be found on the website
{\texttt{http://egg.astro.cornell.edu/alfalfa/}} and references
therein.

This is the first of two papers in which we deal with a compilation of
the new \alf\ data, partially released already in several installments
\citep*{Gio07,Sai08,Ken08,Sti09,Mar09}. The data are from two separate
sky strips of the northern Galactic hemisphere covering a total area
of 9 hr in right ascension and 16 deg in declination. This portion of
the survey is especially interesting because it overlaps almost
entirely with the SDSS footprint, thus offering the possibility of
cross-correlating observations of the cold-gas component with
multi-wavelength optical measurements. This synergy between the
surveys will be exploited in this paper and the accompanying one
\citep[][hereafter Paper II]{Tor10} in order to carry out a systematic
analysis of the fundamental properties of nearby gaseous galaxies and
their interrelations. In this way we hope both to gain a more
complete understanding of the structure of the baryonic component in
these systems and to set up unbiased benchmarks for the normalcy of
their \hi\ content that rely on integral attributes that are easily
accessible to observation. While this is not the first attempt to
conduct this sort of research
\citep[e.g.][]{HG84,RH94,GPB96,SGH96,Gar-A09}, both the significant
increase in the sample size and the high degrees of homogeneity and
completeness of the data will allow us to extend the results of
previous studies by addressing issues that fell outside their scope.

In this paper, we describe the main statistical properties of the two
large catalogs used in our research. We also: describe the steps
followed to cross-correlate the radio and optical sources and to
calculate their radial velocities (Section~\ref{catalogs}); explore
different characterizations of the environment of \alf\ detections
that may be useful for defining suitable reference samples for the
\hi\ content of galaxies (Section~\ref{control_samples}); and carry
out a comparative study of the distribution of the optical structure,
colors, and morphologies of the objects included in the selected
datasets (Section~\ref{optical_properties}). In Paper II, we will
apply strategies of multivariate data analysis to a complete subset of
the \hi\ emitters in low density environments defined here. Our aim in
that paper will be both to explore inter-variable correlations and to
determine the linear combinations of intrinsic properties that best
represent these systems. The relationships we derive will define the
standards of normalcy for the neutral gas mass content of field
galaxies, and thus provide a fundamental reference for future research
into the evolutionary effects of the galactic environment.

Unless otherwise stated, to calculate intrinsic quantities we adopt a
reduced Hubble constant $h=H_0/(100$\,\kms$\,$Mpc$^{-1})=0.7$.

\section{MAIN 21-CM AND OPTICAL CATALOGS}\label{catalogs}

We describe here the primary databases from which we assembled the
galaxy samples used in the present study and examine their statistical
properties. This section also outlines the steps followed to add
information on the radial distance of the selected galaxies to the
measurements already available.

\subsection{Statistical Properties of Catalogs and
  Cross-identification of Sources}\label{cross-correlation}

The main catalog used in this paper contains \alf\ \hi\ measurements
from two regions of the high Galactic latitude sky:
$07^{\mathrm{h}}30^{\mathrm{m}} < \mbox{R.A.}(\mbox{J}2000.0) <
16^{\mathrm{h}}30^{\mathrm{m}}$, $+4\degr <
\mbox{Decl.}(\mbox{J}2000.0) < +16\degr$ and $+24\degr <
\mbox{Decl.}(\mbox{J}2000.0) < +28\degr$. This gives a total coverage
of 2160 deg$^2$.

The \alf\ \hi\ detection threshold and completeness limit depend on
the source signal-to-noise ratio, $S/N$, given by the expression
\citetext{cf.\ \citealt*{Mar09}}:
\begin{equation}\label{SN}
S/N=\frac{1000\,\Fhi}{\W}  \: \frac{w_{\rm smo}^{1/2}}{\sigma_{\rm rms}}\;, 
\end{equation}
where $\Fhi$ is the 21-cm line flux integral in Jy \kms, which in
combination with the cosmological distance to the source, $d$, in
$h^{-1}$~Mpc (see Section~\ref{distances}), provides its total neutral
gas mass in $h^{-2}\,\msun$ units
\begin{equation}\label{MHI}
\Mhi=2.356\times 10^5 d^2 \Fhi\;,
\end{equation}
$\W$ is the observed velocity width of the source line
profile at the $50\%$ level of the two peaks corrected for
instrumental broadening in \kms, $w_{\rm smo}$ is a smoothing width (either
$\W/20$ for $\W< 400$ \kms\ or $20$ for $\W\ge 400$ \kms), and
$\sigma_{\rm rms}$ is the rms noise figure measured in mJy at a resolution of 10 \kms. \alf\ detections are coded according to a scheme dependent
on $S/N$ and availability of a prior, e.g., an optical source that matches
in angular coordinates or redshift. Code 1 sources are detections with 
a $S/N>6.5$ and thought to be reliable with high confidence level; code
2 sources typically have $4<S/N<6.5$, but the marginal \hi\ detection
is corroborated by a match in position and redshift of an optical
source. By searching through the \alf\ data for code 1 and code 2 sources, we came up with $15,043$ extragalactic objects selected purely on
their \hi\ content in the sky region defined above and over the full
survey bandwidth.

\begin{figure}
\begin{center}
\epsscale{0.95}
\plotone{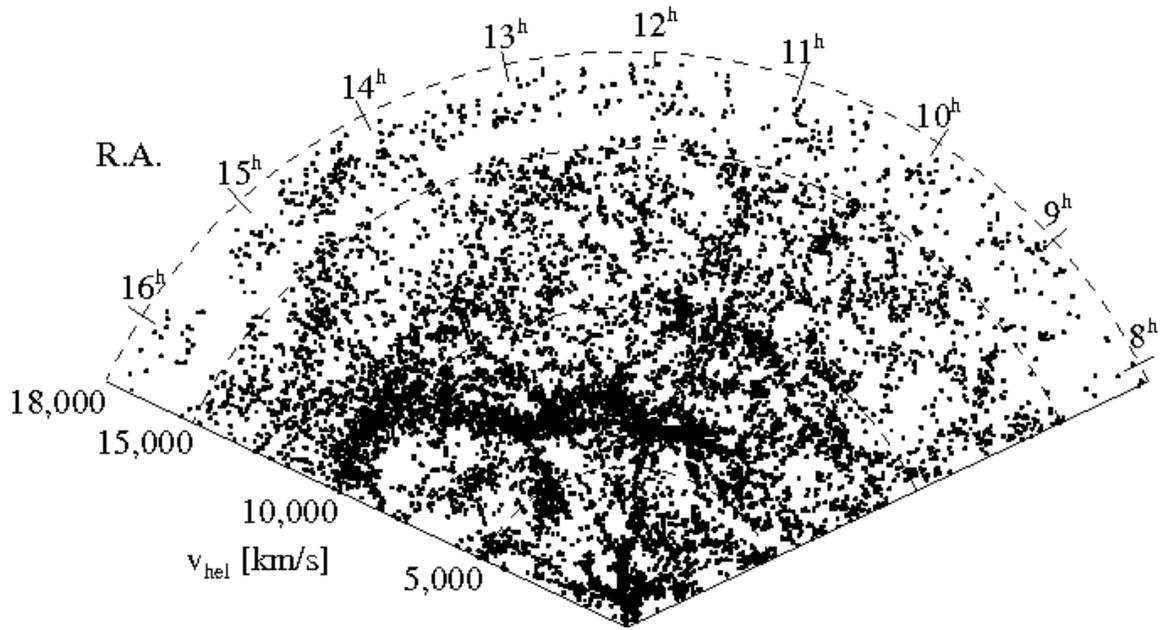} 
\end{center}
\caption{\small Sky distribution of \alf\ sources in the sky region
  where it overlaps with the SDSS DR7 spectroscopic
  survey.} \label{fig_wedge_all}
\end{figure}

The portion of the \alf\ survey we use overlaps with other major
galaxy surveys such as 2MASS and SDSS. The latter ---which has scanned
about one quarter of the northern Galactic cap sky in the bands $u$,
$g$, $r$, $i$, and $z$, reaching objects up to a total magnitude of 22.2
in $r$ band \citep{Sto02}--- is especially interesting for our
purposes, as it provides a uniform, very extensive list of photometric
and spectroscopic measurements suitable for carrying out an exhaustive
study of the correlations between the gaseous and stellar
properties of galaxies. Here we make use of SDSS data coming from the
full survey \citep[SDSS DR7;][]{Aba09}, which also provides radial
velocity measurements for a considerable number of its main
targets. The spectroscopic observations, which are complete to a
Petrosian $r$-band magnitude of 17.77 after correcting for Galactic
extinction (but not for internal reddening), offer a redshift coverage
that is deep enough to allow the identification of the most
probable optical counterpart to all \alf\ detections found out to
$\sim 260$~Mpc within the sky region where the two surveys
overlap\footnote{See
  \texttt{http://egg.astro.cornell.edu/alfalfa/images/alfalfa\_sdss\_coverage.jpg}
  for an image of the footprint of \alf\ on the SDSS, whose geometry
  we defined using mangle polygons \citep{HT04}.}. A graphical
representation of the positions of the \alf\ sources in the
overlapping region is provided in Figure~\ref{fig_wedge_all} by means
of a wedge diagram in right ascension. The total number of \hi\
detections coded 1 and 2 in the SDSS overlap region is 
$11,239$ after discarding objects that lie closer than 1.5 deg to M87, where
\hi\ sources cannot be reliably detected because of the strong radio
continuum emission.

\begin{figure}

\begin{center}

\epsscale{1}
\plottwo{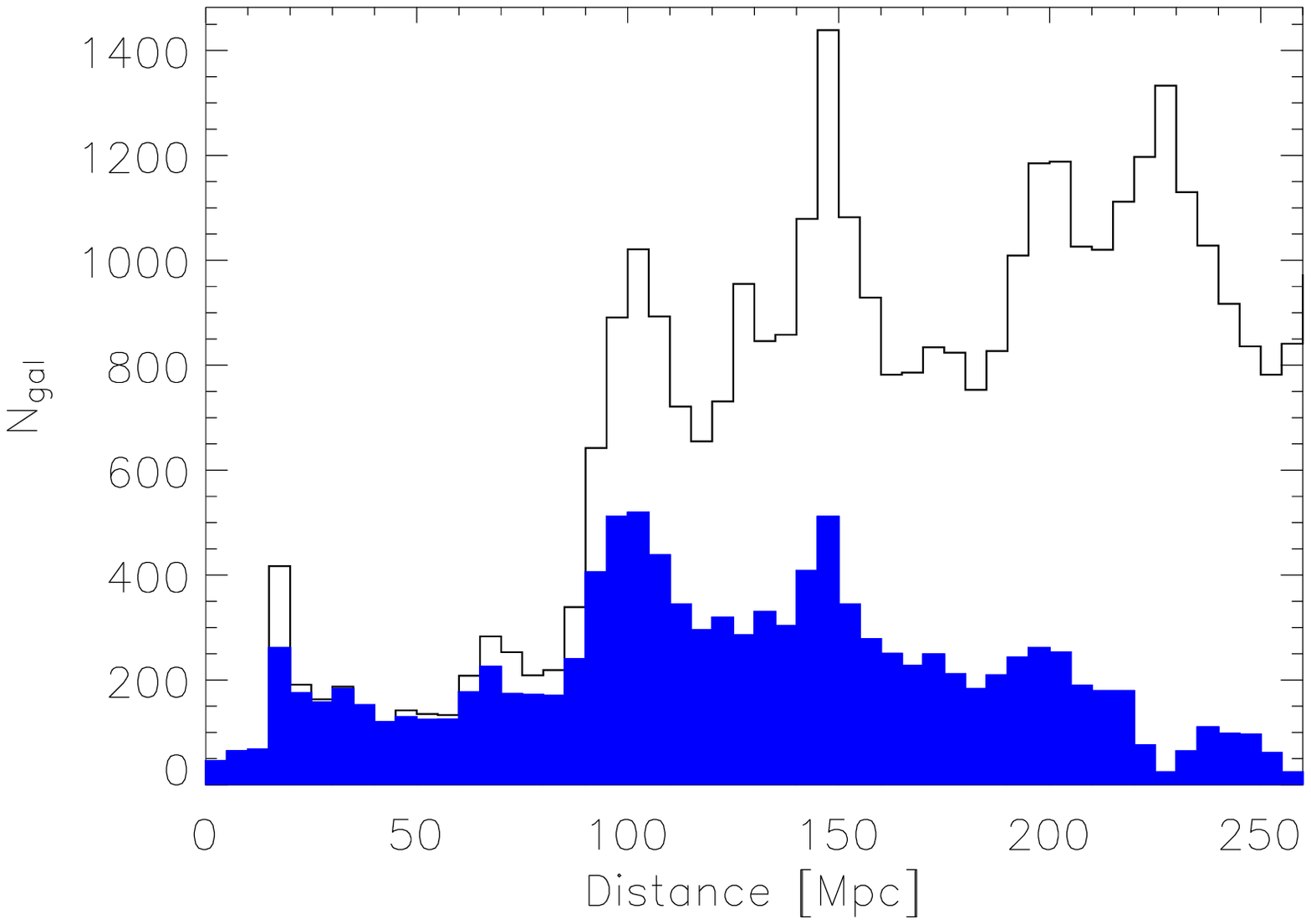}{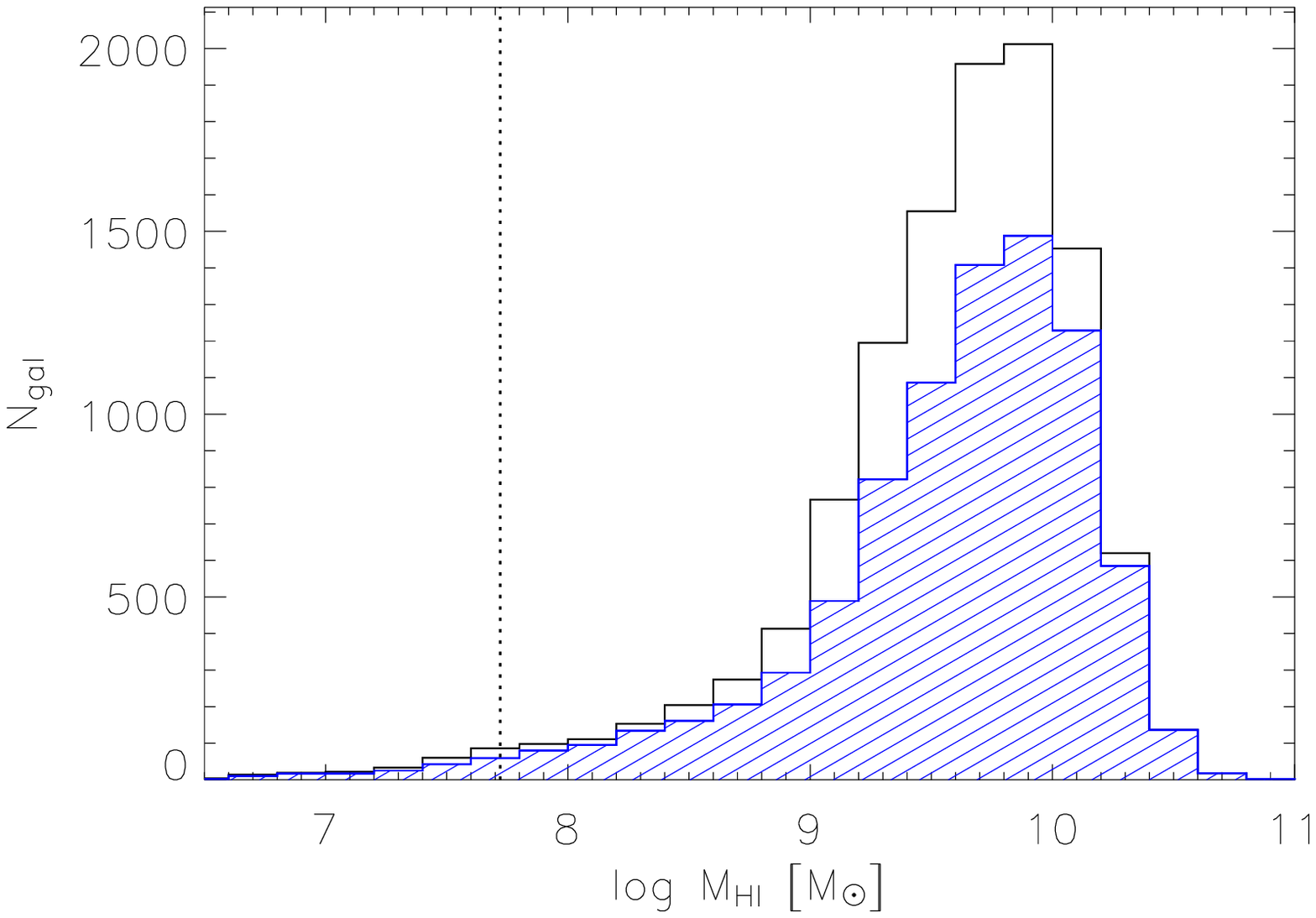}
\plottwo{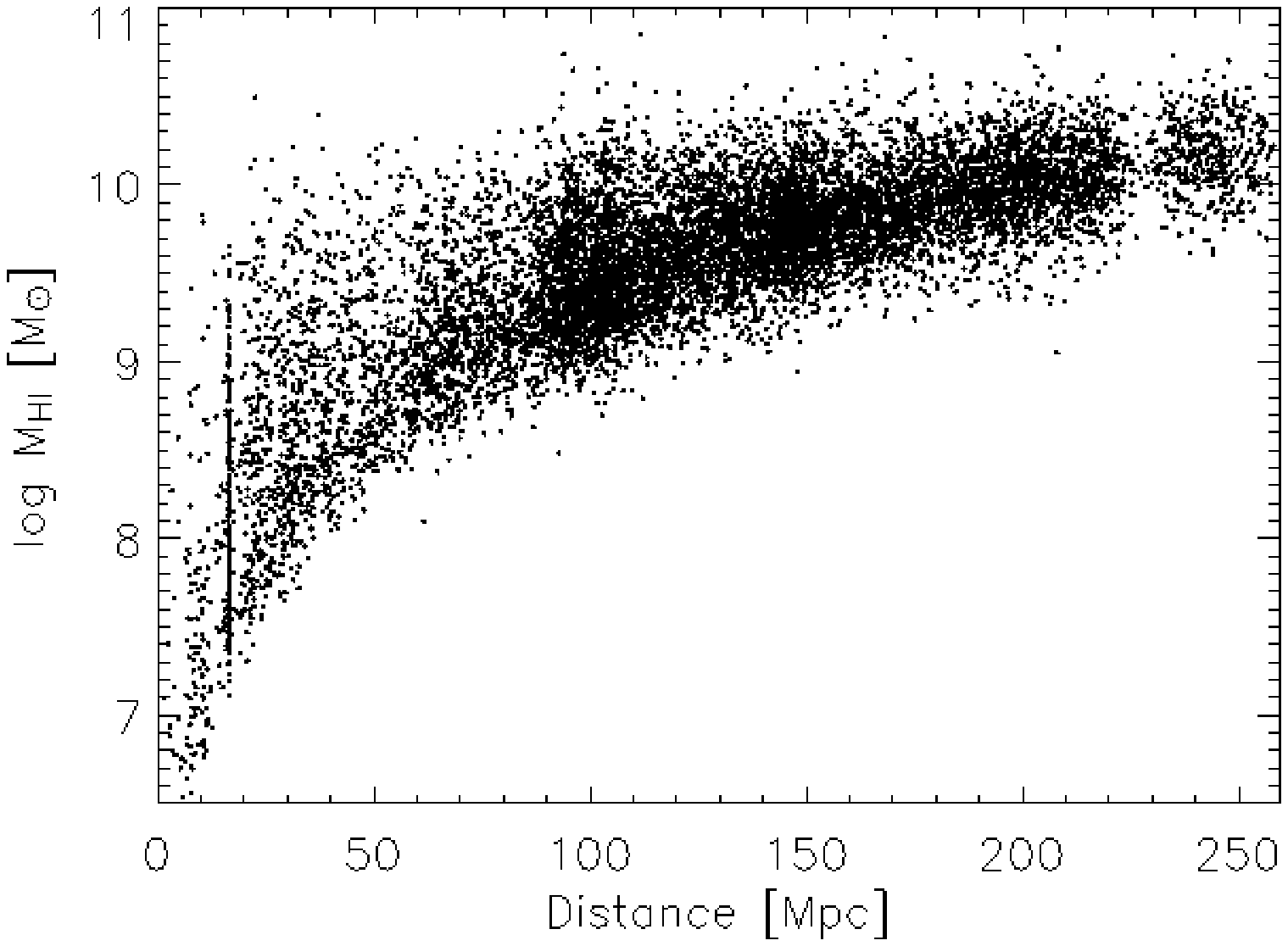}{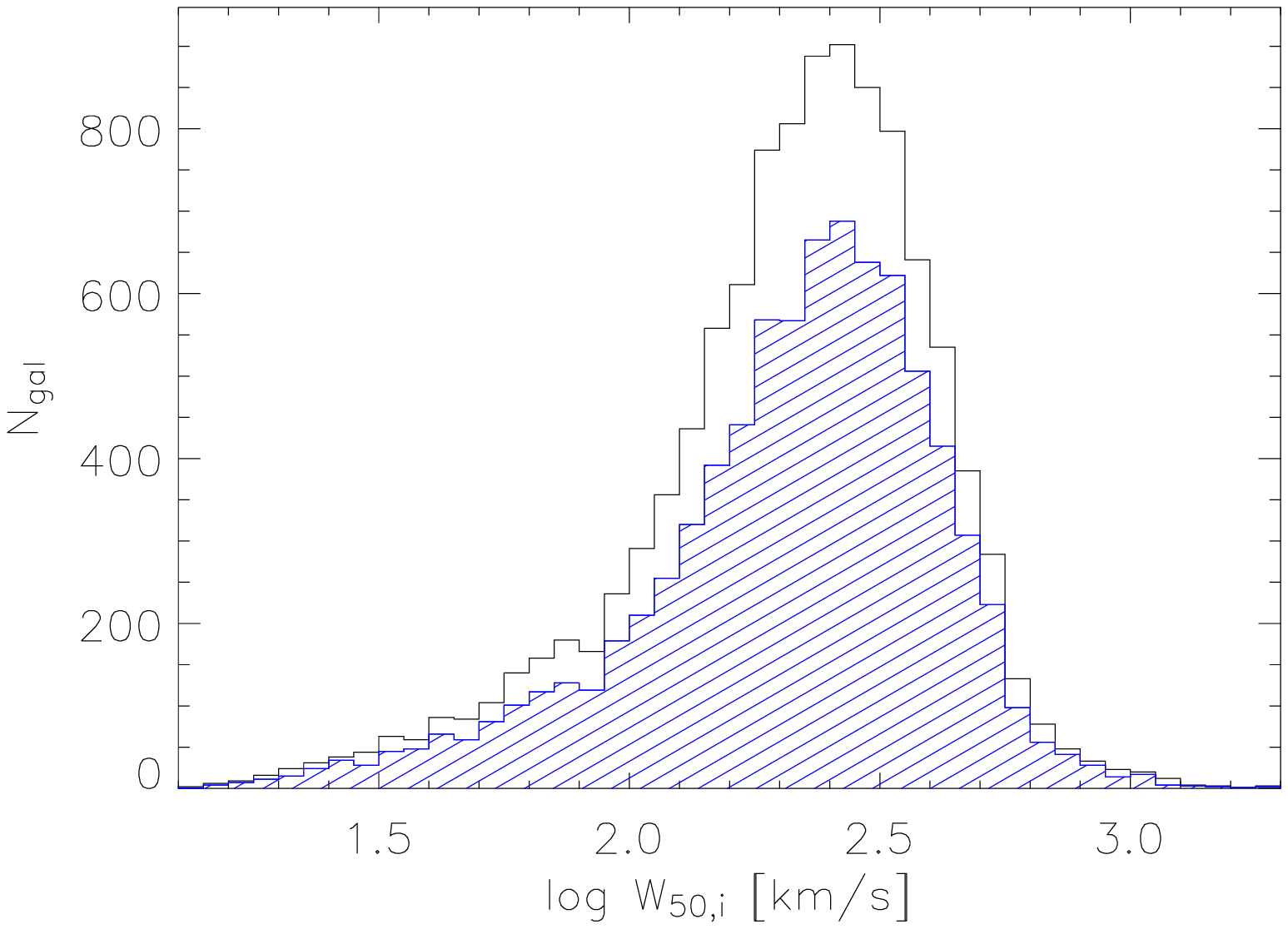}

\caption{Top-left: Distribution of \alf\ (solid-blue histogram) and
  SDSS (black line) galaxies according to CMB distance (see
  Section~\ref{distances}). Top-right: \hi\ mass distribution for
  \alf\ detections (solid line). The distribution for the \alf\ code 1
  detections is represented by the dashed-blue histogram.  The
  vertical dotted line corresponds to the minimum \hi\ mass detected
  at 2000 \kms. Bottom-left: Sp\"anhauer diagram for the objects in
  the \alf\ catalog. Bottom-right: Inclination-corrected velocity
  width distribution (black line) for \alf\ sources. The dashed-blue
  histogram depicts the distribution arising from code 1 objects
  only.}\label{fig_properties}

\end{center}
\end{figure}

Some of the main statistical properties of this latter data set are
displayed in Figure~\ref{fig_properties}. In the top left panel, we
show the distance distribution of the \alf\ sources. The most
important large-scale features are reflected in the various peaks
between $\sim 80$ and $160$~Mpc (see also
Figure~\ref{fig_wedge_all}). The shape of this histogram follows that
produced by the distance distribution of SDSS galaxies (black solid
line) closely, except for the narrow artificial dip seen near $\sim 125$~Mpc
and the broader one between $\sim 215$ and $230$~Mpc. Both of these
are related to the important loss in sensitivity of the \alf\ survey
due to radio frequency interference (RFI) coming from the San Juan
airport's radar \citep{Gio07}. The \hi\ mass distributions for the
full data set and from code 1 detections only (hatched histogram) are
shown in the top right panel of Figure~\ref{fig_properties}. The
vertical dotted line illustrates the minimum
detectable \hi\ mass arising from the lower heliocentric velocity
cutoff of 2000 \kms\ adopted in Section~\ref{local_density} so as not to include 
the most uncertain values of this quantity. (Note that this also leaves
many nearby dwarfs that are rich in neutral
gas out of the present study.) Below, the bottom left panel features the Sp\"anhauer diagram,
detailing the \hi\ mass vs.\ CMB distance. The broad gap in detections
associated with the region most affected by the intrusion of RFI is
clearly visible near the right-hand end. Finally, the bottom right panel of
Figure~\ref{fig_properties} shows the intrinsic velocity width
distribution. As in the top right panel, we also depict the
distribution arising only from code 1 objects to demonstrate that the
inclusion of code 2 detections does not bias the results.

\begin{figure}
\epsscale{0.7}
\plotone{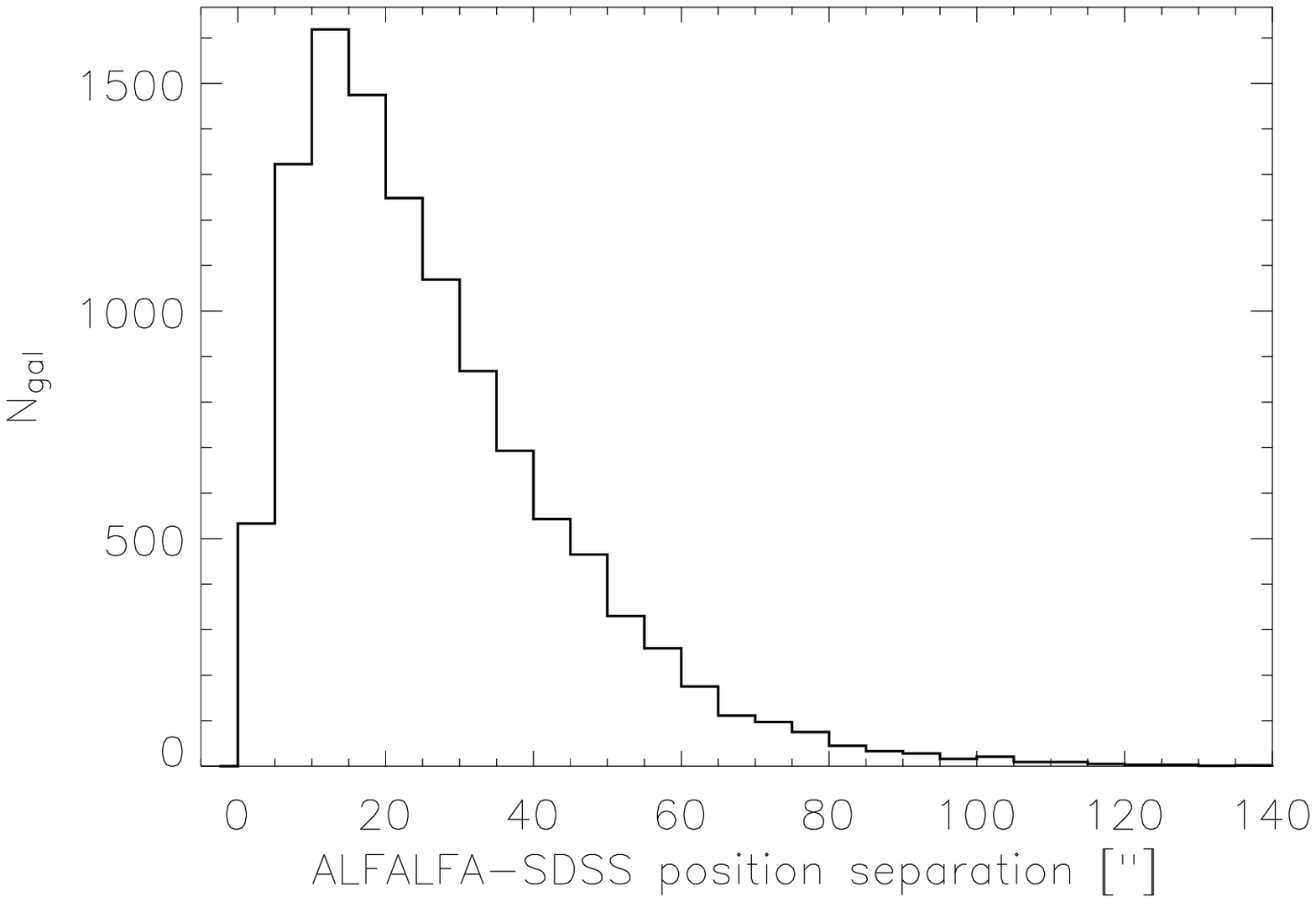}
\plotone{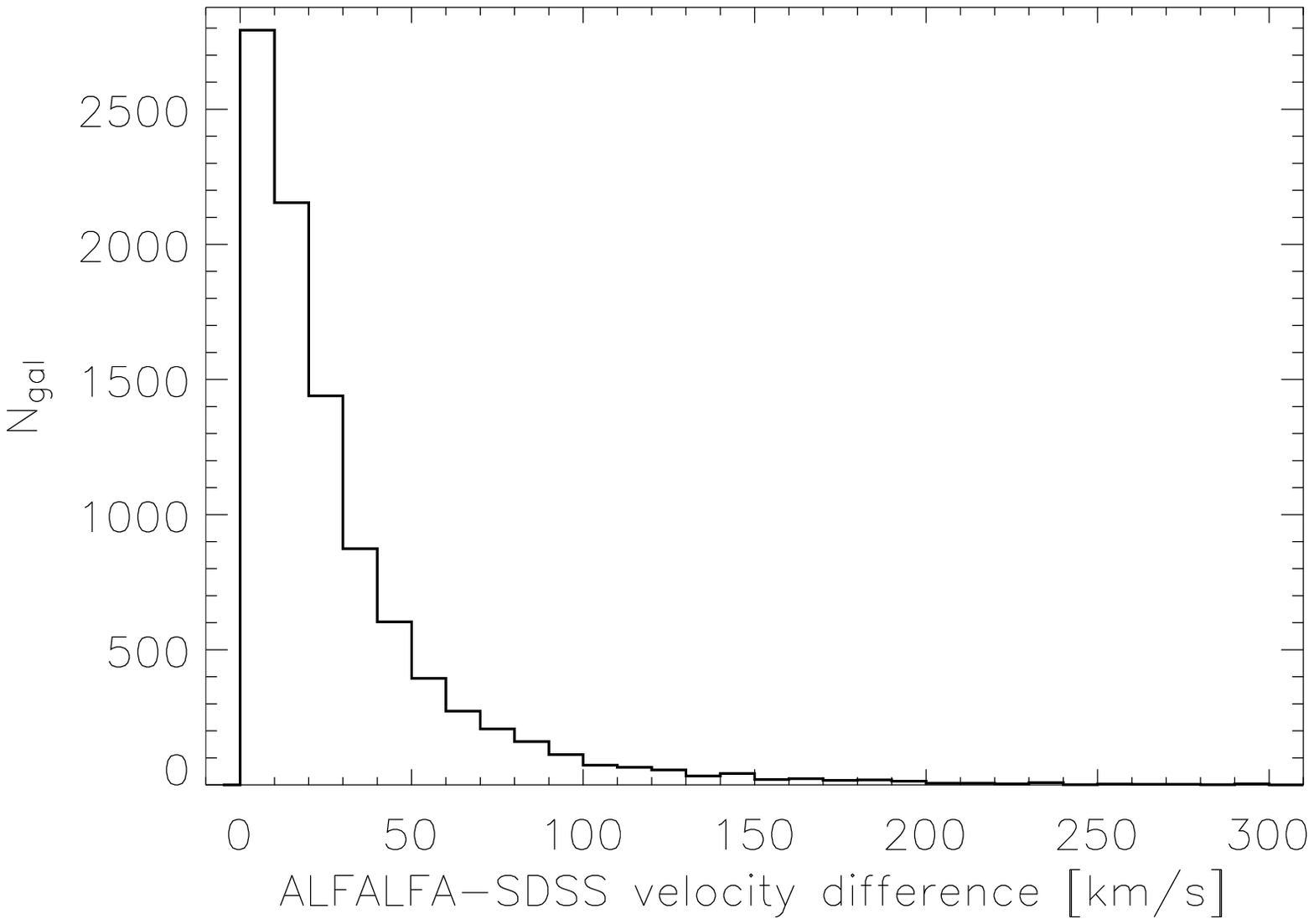}
\caption{\small Top: Distribution of differences between \hi\ and SDSS
  sky positions for \alf\ detections with a SDSS photometric
  counterpart. Bottom: Distribution of differences between \hi\ and
  SDSS radial heliocentric velocities for \alf\ detections with a SDSS
  spectroscopic counterpart.} \label{fig_match_sdss}
\end{figure}

Unlike other \hi\ surveys, the improved angular resolution of \alf\
---which leads to positional accuracies typically better than 20
arcsec \citep{Gio07}--- usually makes the identification of the optical
counterpart of a given \hi\ source an unambiguous
task. Furthermore, the fact that, for each \hi\ detection, the \alf\
catalog provides the centroid of the optical galaxy that appears to be
the most plausible counterpart according to proximity in $z$-space,
color, and morphology \citep{Gio07}, enabled us to perform a
semi-automated catalog cross-correlation. Specifically, we
assumed that the optical counterpart of a \hi\ detection provided
by the \alf\ catalog and a spectroscopic SDSS detection are the same
object if their projected separation on the plane of the sky is not
larger than the size of the optical source and the difference in the
reported velocities is less than 300 \kms. We allow for such a
relatively large discrepancy between the observed heliocentric
velocities in order to account for the fact that SDSS spectroscopic
measurements on an extended object do not always reflect the radial
velocity of the object's centroid, so a positive match between two
galaxies relies essentially on the similarity of their sky
coordinates. In practice, however, the dispersion in the difference
between radial velocities of matched pairs is only of $\sim 35$ \kms\
(see Figure~\ref{fig_match_sdss}). For the cases in which \alf\
sources were lacking a spectroscopic counterpart, the photometric
criterion was applied. Human intervention was reduced to only cases of doubtful
identification.

Table~\ref{tab_alfalfa_detection_frac} lists the numbers (and
fractions) of \alf\ objects with a photometric or a spectroscopic
counterpart in the SDSS for different bins of radial heliocentric
velocity. Only $\sim 2\%$ of the \hi\ sources are not associated with
galaxies listed in the SDSS photometric catalog, while there is a
somewhat larger percentage, $\sim 15\%$, of \hi\ detections without a
spectroscopic counterpart. Note also how the slight incompleteness of
the SDSS DR7 data set that affects mainly the galaxies with the largest
apparent brightness \citep[e.g.,][]{Str02} leaves its imprint in the
form of reduced fractions for the first two velocity bins of the third
and fourth columns of the table. Also included in
Table~\ref{tab_alfalfa_detection_frac} are the numbers (and fractions)
of \alf\ objects detected within the 'SDSS spectroscopic galaxy
sample', SDSS-spec sample for short. From now on, we will use this
shorter designation to refer to the subset of SDSS sources with both
photometric and spectral morphology classified as of galaxy type, and
with $r < 17.77$ mag. About one in every five galaxies in this
subset is also an \alf\ detection, although the fraction rises to one
in two for $cz_{\rm{hel}}\leq 7500$ \kms. The small value quoted in
the higher velocity entry of the last column of
Table~\ref{tab_alfalfa_detection_frac} also reflects the impact of the
substantial incompleteness of the \alf\ data due to RFI above $\sim
15,000$ \kms. The effects on the completeness of a few other narrower
RFI features present in the spectral bandpass are comparatively
modest.

\begin{figure}
\epsscale{0.8}
\plotone{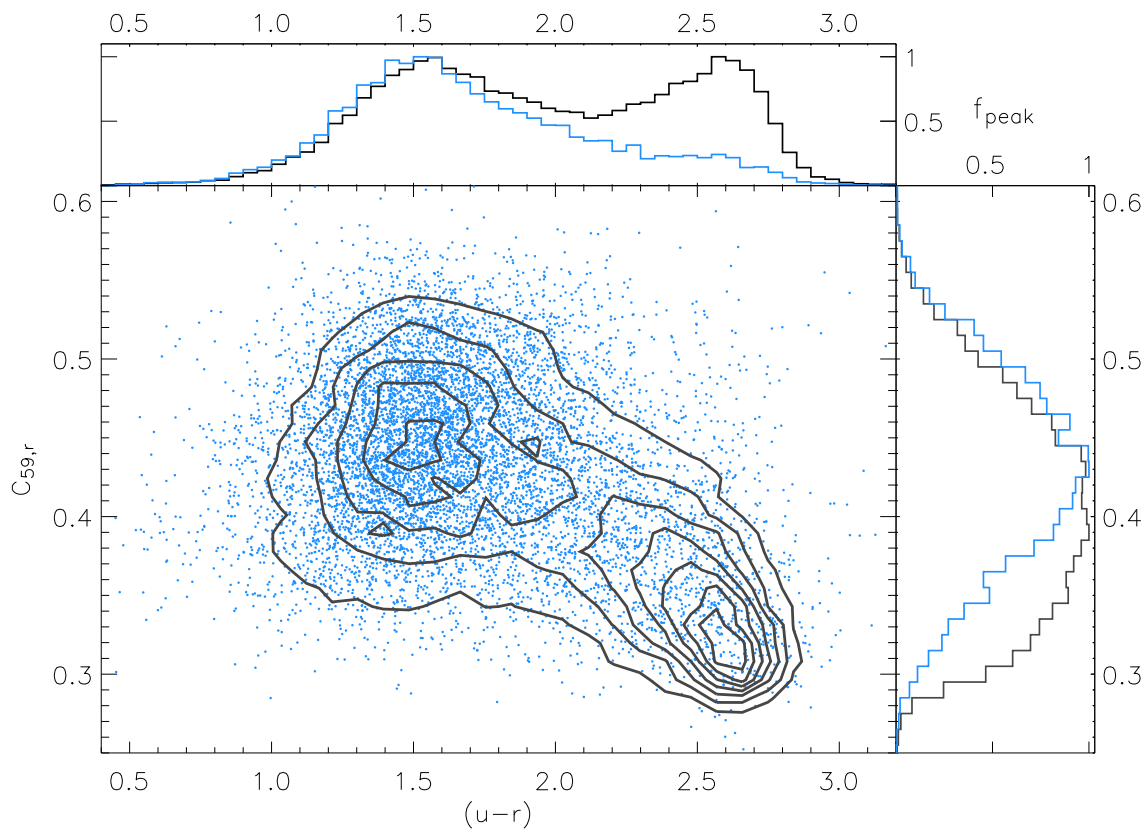}
\vskip0.25in
\caption{\small Inverse light concentration index in the $r$ band,
  $\rici$, versus $(u-r)$ color for members of the SDSS spectroscopic
  sample (black contours), and for those that are also \alf\
  detections (blue dots). Contours are spaced at intervals of 40
  density units. The histograms on the right and top sides of the main
  panel show the distributions of the corresponding variables for
  these two data sets normalized to peak
  values.}\label{fig_color_ci_diagram}

\end{figure}

Figure~\ref{fig_color_ci_diagram} shows how the galaxies
in both catalogs are distributed in a global color-light concentration
parameter space. The plot shows $(u-r)$ color versus inverse light
concentration index, defined as the ratio of the radii containing
$50\%$ and $90\%$ of the Petrosian galaxy light, $\ici=R50/R90$, which
we calculate in the $r$-band. The density contours delineate the
behavior of the galaxies within the SDSS-spec sample, while the data
points correspond to those objects that are also \alf\ members. It is
clear from this figure that SDSS galaxies show a clear bimodality
and separate into two main classes with comparable numbers of objects
that can be addressed as the blue- and red-sequence galaxies
\citep[e.g.][]{Bla03}. In contrast, \hi\ detections mostly populate 
the locus corresponding to the bluer, late-type, star-forming systems.
This trend has been known about for a very long time \citep[e.g.,][]{RH94}
and it is in very good agreement with the similar offset found by
\citet{Wes09} using 195 \hi-selected galaxies from the Parkes
Equatorial Survey \citep{Gar-A09}. We will show in
Section~\ref{properties_LDE} that, while the bimodality of the
optically-selected sources is reinforced when looking at the global
trend with increasing radial velocity, and hence with increasing
typical luminosity of the galaxies, the \hi\ detections are distributed in a
similar way independently of their distance from the observer.

\subsection{Velocity and Radial Distance of
  Sources Referred to the CMB Frame}\label{distances}

To calculate radial velocities and distances referred to the Cosmic
Microwave Background (CMB) standard of rest for the galaxies in both
the \alf\ and SDSS spectroscopic data sets, we first correct the
heliocentric velocities listed in these two catalogs for peculiar
motions. This is done by adopting the multiattractor model of the
local peculiar velocity field determined by \citet{Mas05} from the
best fit to data from the new $I$-band Tully-Fisher catalog SFI++
\citep{Spr07}. This model includes the spherical infall into Virgo and
the Hydra-Centaurus supercluster, a quadrupole correction to account
for the smaller than average expansion out of the supergalactic plane,
as well as a residual dipole pointing towards the
Hydra-Centaurus/Shapley superclusters. The predicted peculiar motion
corrections, typically of the order of 200 \kms, are non-negligible
for objects within 6000 \kms. Beyond that radial velocity we assume
the Hubble law is exact. The peculiar velocities are tapered to zero
between 5000 and 6000 \kms\ in order to provide a smooth transition
between these two regimes. For transformations between heliocentric
and CMB frames of reference, we apply the corrections quoted in
\citet{Kog93}. For conversion to a CMB luminosity distance, we
assume Euclidean properties of space, an approximation that is
maintained for the calculation of any intrinsic galactic property
involved in the present study.

To minimize the impact of redshift-space distortions in the assignment
of distances (and, consequently, in local density estimates; see
Section~\ref{local_density}) to galaxies lying in the vicinity of the
groups and clusters present in the cosmic volume targeted by \alf, we
adopt a strategy that can easily be applied in a systematic way to any
virialized aggregation of galaxies independently of its scale. It
consists of estimating the virial radius of the system,
$r_{\mathrm{vir}}$, and its velocity dispersion along the
line-of-sight, $\sigma$, and then using this information to assign the
corresponding system distance to all galaxies with \emph{projected}
distances to the center of one of these overdense environments of less
than $1.5\,r_{\mathrm{vir}}$ and systemic velocities with respect to
the system mean velocity less than $3\,\sigma$. To calculate
$r_{\mathrm{vir}}$ in Mpc, we use the following approximation to the
halo density-radius relation based on a singular isothermal sphere
with mean interior density 200 times the current value of the critical
density of the universe (cf.\ \citealt*{CYE97}):
\begin{equation}
r_{\rm{vir}}\simeq\frac{\sqrt{3}\,\sigma}{700}\;.
\end{equation}
Values for $\sigma$, in \kms, mostly come from the compilation of Abell cluster
properties by \citet{SR91}, while for the few galaxy aggregations
whose velocity dispersion is not quoted in that catalog, we use
the membership assignments carried out by \cite{Spr07} in the standard
estimator \citep[e.g.,][]{Tag08}
\begin{equation}
 \sigma^2=\frac{1}{(1+\bar{z})^2}\sum_{i=1}^{N_v} \frac{(cz_i-c\bar{z})^2}{N_v-1}\;,
\end{equation}
where the sum is intended over the $N_v$ members of the aggregation
with measured velocity $cz_i$, and $\bar{z}$ is the mean system redshift.

\section{\hi\ CONTROL SAMPLES}\label{control_samples}

Because the cold gas component of galaxies is both a good indicator of
their potential for star formation and an excellent probe of the
physical conditions of the intergalactic medium, the quantification of
the \hi\ content in galaxies is a useful diagnostic tool for studying
the evolutionary effects of their surroundings.

The definition of a benchmark suitable for absolute measures of \hi\
content necessitates a control sample as bias-free as possible formed
by galaxies expected to show little or no evidence of environmental
interactions. In this section, we study different
characterizations of the local environment of \alf\ detections, define
three possible control samples (see Table~\ref{tab_samples}), and
explore their relative degree of adequacy.

\subsection{Local Number Density Measurements for \alf\ and SDSS
  Galaxies}\label{local_density}

The environment of a galaxy can be defined in terms of the density of
galaxies located in its immediate vicinity. To calculate the
underlying three-dimensional galaxy number density at a given
position, we use the galaxies in the SDSS-spec data set as the nodes
of a $n$th-nearest neighbor network. This calculation is carried out
for both \alf\ and SDSS sources, excluding galaxies lying in those
regions in which the results are affected by large errors or where
detections are very scarce and/or unreliable. In particular, we impose
a lower heliocentric velocity limit of 2000 \kms\ to avoid those
galaxies likely having the largest relative uncertainty in distance
arising from peculiar motions not duly accounted for ---a constraint
that also removes the closest cosmic volume where the SDSS pipeline
photometry of large, extended galaxies is most inaccurate---, and a
velocity cutoff ranging from 15,000 \kms\ up to 16,000 \kms\ that
embraces the deep trough in sensitivity of the \alf's spectral window
due to RFI (see Figures~\ref{fig_wedge_all} and \ref{fig_properties}).

We follow \citet{CH85} and compute a local, continuous measure of
environment from the unbiased estimator
\begin{equation}\label{rho_6} 
\rho_6=\frac{15}{4\pi d_6^3}\;,
\end{equation} 
where $d_6$ is the three-dimensional (comoving) distance to the
6th-nearest neighbor in the SDSS-spec sample in $h^{-1}$ Mpc. In an
effort to correct for redshift-space distortions\footnote{We do not
  correct for survey edge effects given the large size of the
  catalogs.} and improve the accuracy of our density measures in
overdense environments, $\rho_6$ is calculated, for those galaxies
assigned to a cluster (Section~\ref{distances}), using instead
projected separations scaled by a factor of $(3/2)^{1/2}$ to convert
them into a three-dimensional measure of distance \citep*{Coo05}.

Local densities calculated this way are then corrected for the
magnitude cutoff of the SDSS catalog by multiplying them by the
inverse of the selection function
\begin{equation}\label{sel_func}
\psi(\alpha,\delta,z)=\frac{ \int^{M_{\mathrm{lim}}(z)+\Delta
M(\alpha,\delta)}_{-\infty}\phi(M)\mbox{d}M}{\int^{M_\mathrm{fid}}_{-\infty}\phi(M)\mbox{d}M
}\;,
\end{equation}
which we calculate by integrating the differential luminosity
function, $\phi(M)$, of the SDSS provided by \citet{Bla01}. In
equation~(\ref{sel_func}), $M_{\mathrm{lim}}(z)$ and
$M_{\mathrm{fid}}\equiv M_{\mathrm{lim}}(z_{\mathrm{fid}})$ are,
respectively, the faintest magnitudes detectable because of the
optical flux limitation at the galaxy redshift and at a fiducial
redshift, $z_{\rm{fid}}$, which we match up with that corresponding to
the low velocity cutoff of 2000 \kms\ imposed to the samples, while
$\Delta M(\alpha,\delta)$ is the extinction correction as a function
of right ascension ($\alpha$) and declination ($\delta$), for which we
use the \citet*{SFD98} values loaded in the SDSS database.

It is worth noting, however, that this measure of the local density
cannot be expanded to the full \alf\ survey, which is also carrying
observations within the solid angle
$22^{\mathrm{h}}\leq\mbox{R.A.}(\mbox{J}2000.0)\leq 03^{\mathrm{h}}$
and $0\degr\leq\mbox{Decl.}(\mbox{J}2000.0)\leq +36\degr$, not
scanned by the SDSS. In this case, the 2MASS \citep{Skr06}, with its
full sky coverage ($98\%$ for the extended source catalog), offers the
most sensible alternative to characterizing the environment in which
\alf\ detections reside. Thus, in anticipation of future research, we
also assigned a \emph{surface} density parameter $\mu_6$ to each
optical and gaseous spectroscopic source from the formula
\begin{equation}\label{mu_6}
  \mu_6={5\over{\pi s_6^2}}\;,
\end{equation}
with $s_6$ the projected distance to the 6th-nearest neighbor, in
degrees, derived taking into account the underlying 2MASS galaxy 
distribution on the sky in the $K_s$ band down to 13.0 mag, where the
2MASS extended source catalog has been estimated to be at least $95\%$
complete for most of the sky \citep{Jar00}. We do not correct this
surface density estimator for Galactic absorption, because at this
extreme near-infrared wavelength the opacity is insignificant for
$\left| b^{II}\right|\gtrsim 10\degr$. It should not be forgotten,
however, that this immunity to the effects of the Milky Way's dust
arising from 2MASS being a survey in the near-infrared, biases its
sensitivity regarding the morphology of the detected galaxies toward
early-type spirals and ellipticals.

\begin{figure}
\epsscale{0.6}
\plotone{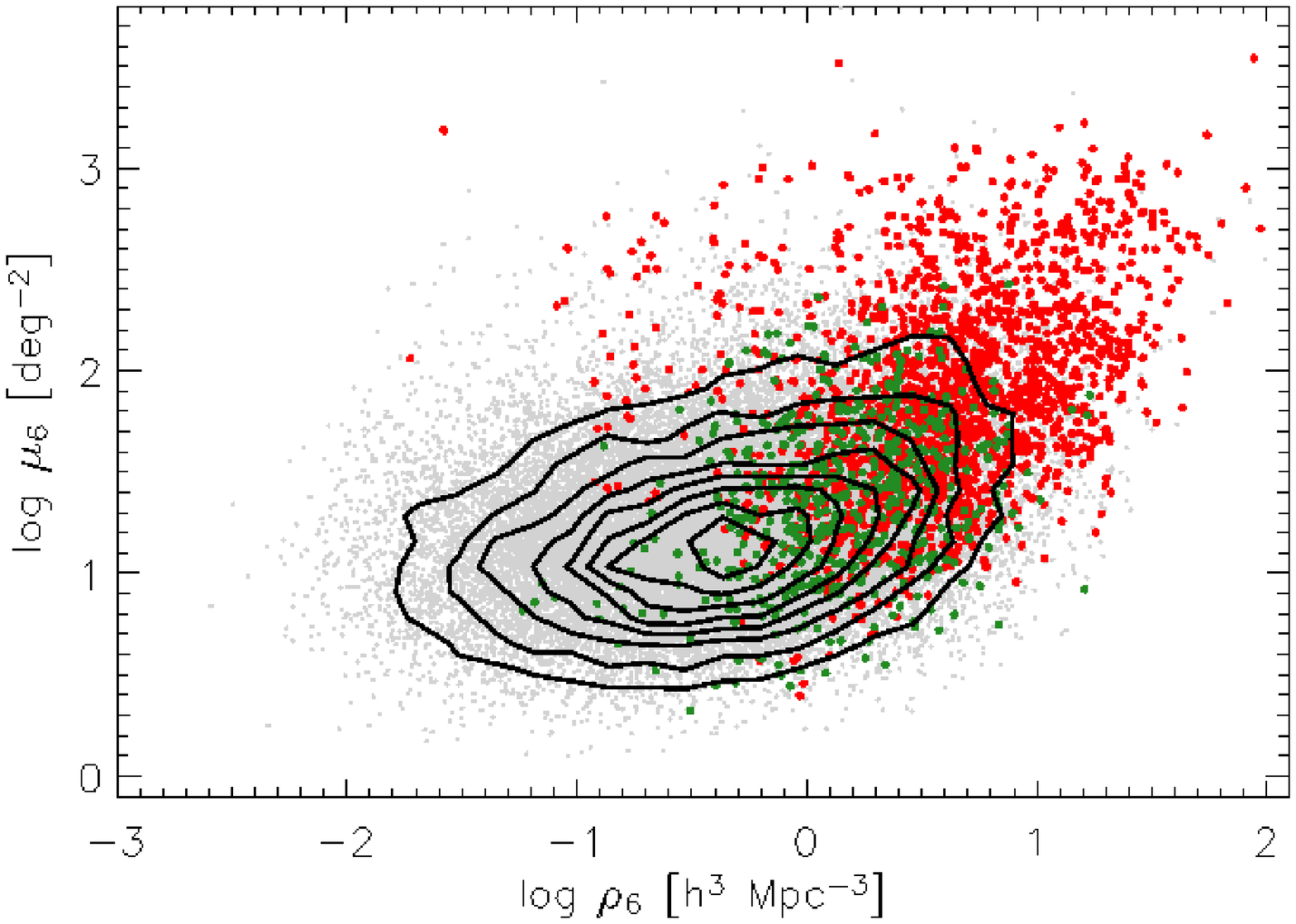}
\plotone{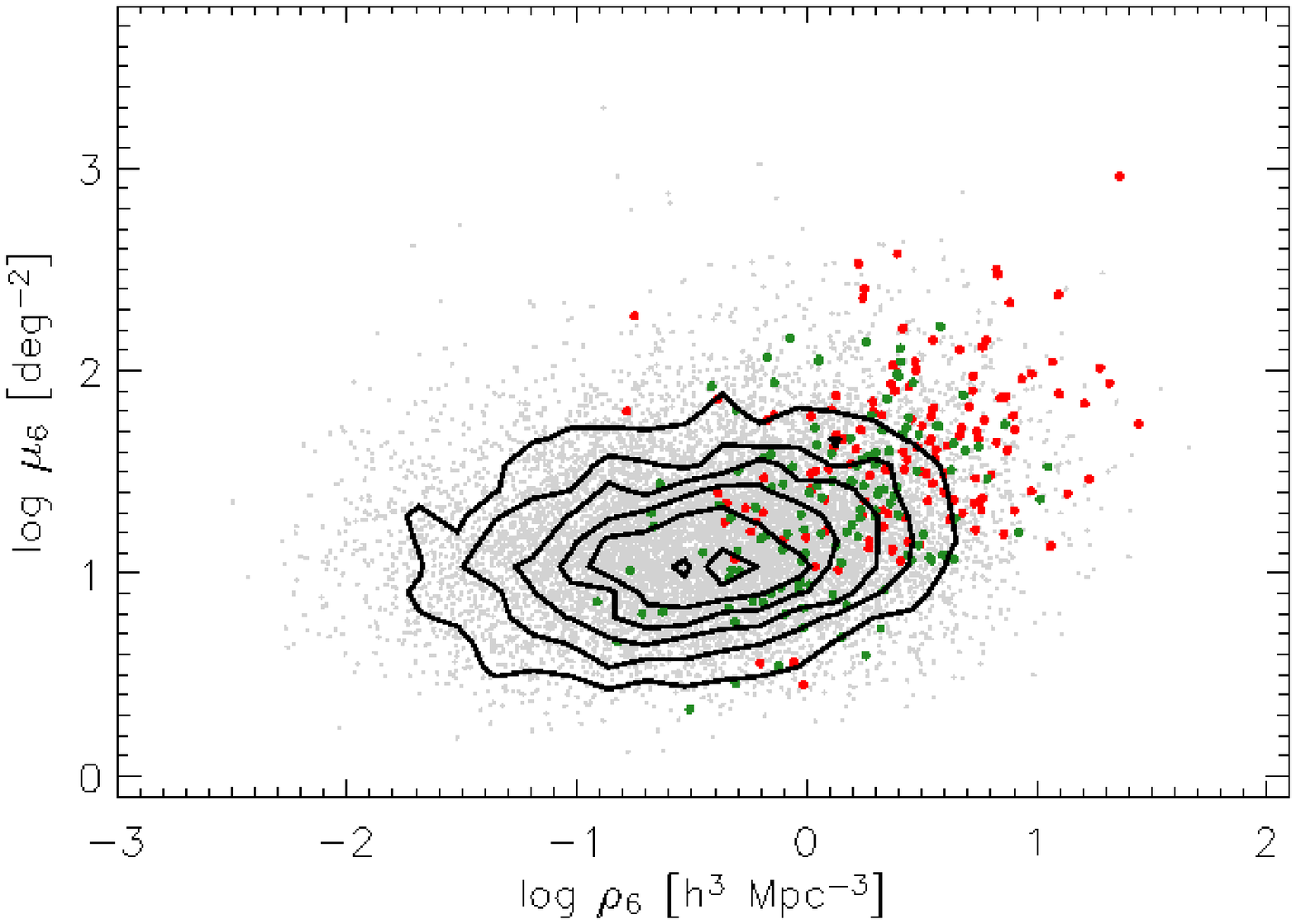}
\caption{\small Three-dimensional local density of galaxies derived
  using the distance to the 6th-nearest neighbor, $\rho_6$
  (Equation~\ref{rho_6}), in the SDSS-spec data set vs.\ the
  corresponding projected density estimator, $\mu_6$
  (Equation~\ref{mu_6}), calculated from the 2MASS galaxy distribution
  in the $K_s$ band down to 13.0 mag. We show results for SDSS-spec
  galaxies (top) and for   \alf\ detections (bottom). Contours are 
  spaced at intervals of 50 (top) and 25 (bottom) density 
  units. Galaxies classified as cluster members are drawn using big 
  red dots, whereas green dots are used for the clusters' outskirts 
  population. In these plots, the values of $\rho_6$ for cluster 
  members have not been corrected for redshift-space 
  distortions.}\label{fig_rhovsmu}

\end{figure}

In Figure~\ref{fig_rhovsmu}, we compare the values of $\rho_6$
(uncorrected for redshift-space distortions) and $\mu_6$ for galaxies
in the SDSS-spec (top) and \alf\ (bottom) data sets. The
plots show that these two local density tracers follow a weak,
non-linear relationship, with the three-dimensional measure of
environment spanning a somewhat larger dynamic range. Note the
relatively few observations associated with clusters in the bottom
plot caused by the undersampling of the highest density environments
by \alf\ (see also next section).

\subsection{The Low Density Environment \hi\ Galaxy Sample}\label{LDE}

One possibility to define a sample of galaxies whose evolution is
hypothetically less influenced by external, disruptive processes is to
select those objects least affected by the clustering
phenomenon. According to observational findings the influence of the
cluster environment on the properties of galaxies can extend up to
about twice the cluster virial radius
\citep[e.g.,][]{Bal98,Sol01}. Therefore, any density threshold adopted
to extract a control sample of unperturbed galaxies from a given
catalog should be low enough to leave out of it, not only bona fide
cluster and group members, but also those objects located in adjacent
regions.

We studied the relationship between the neutral gas content of
\alf\ galaxies and their local three-dimensional number density
measured in the previous section to see if we can identify the density
threshold above which environmental interactions start to become
important. Following the works by \citet{HG84} and \citet{SGH96}, the
former property has been quantified by means of a \hi-deficiency
parameter, $\df$, that compares (the decimal logarithm of) the
observed \hi\ mass, $\Mhi^{\rm obs}$, in solar units, with the value
expected from a galaxy free of external influences of the same
observed morphological type, $T^{\rm obs}$, and optical linear
diameter, $D^{\rm obs}$, expressed in kpc. Specifically:
\begin{equation} \label{defHI}
\df=\;\langle\log \Mhi(T^{\rm obs},D^{\rm obs})\rangle-\log \Mhi^{\rm obs}\;,
\end{equation}  
so positive values of $\df$ indicate \hi\ deficiency. This test has
been restricted to those \alf\ detections with Hubble types and
(apparent) blue visual sizes available on the Arecibo General Catalog
(AGC), a private galaxy database maintained by MPH and RG. In order to
increase sample size and reduce statistical errors, we inferred
expectation values for the \hi\ mass for 2615 disk galaxies with
Sa--Sd morphologies using the maximum likelihood linear regressions of
$\log \Mhi$ on $\log D^{\rm obs}$ given in \citet{SGH96}. Since in
that work the standards of normalcy for the \hi\ content were defined
specifically for galaxies of type Sa--Sc, we thus assigned Scd--Sd
types to be the same as Sc. 

\begin{figure}
\epsscale{1}
\plottwo{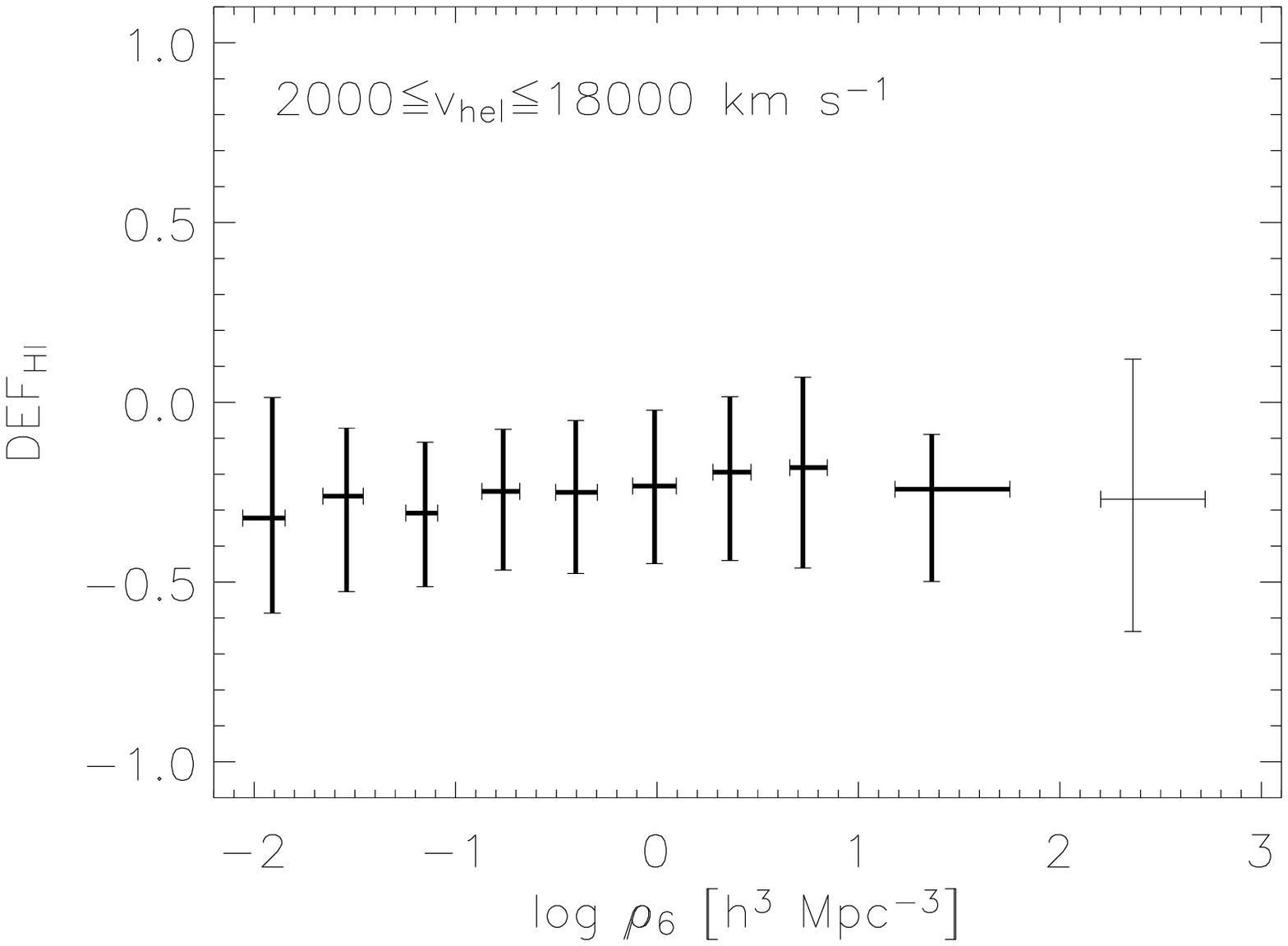}{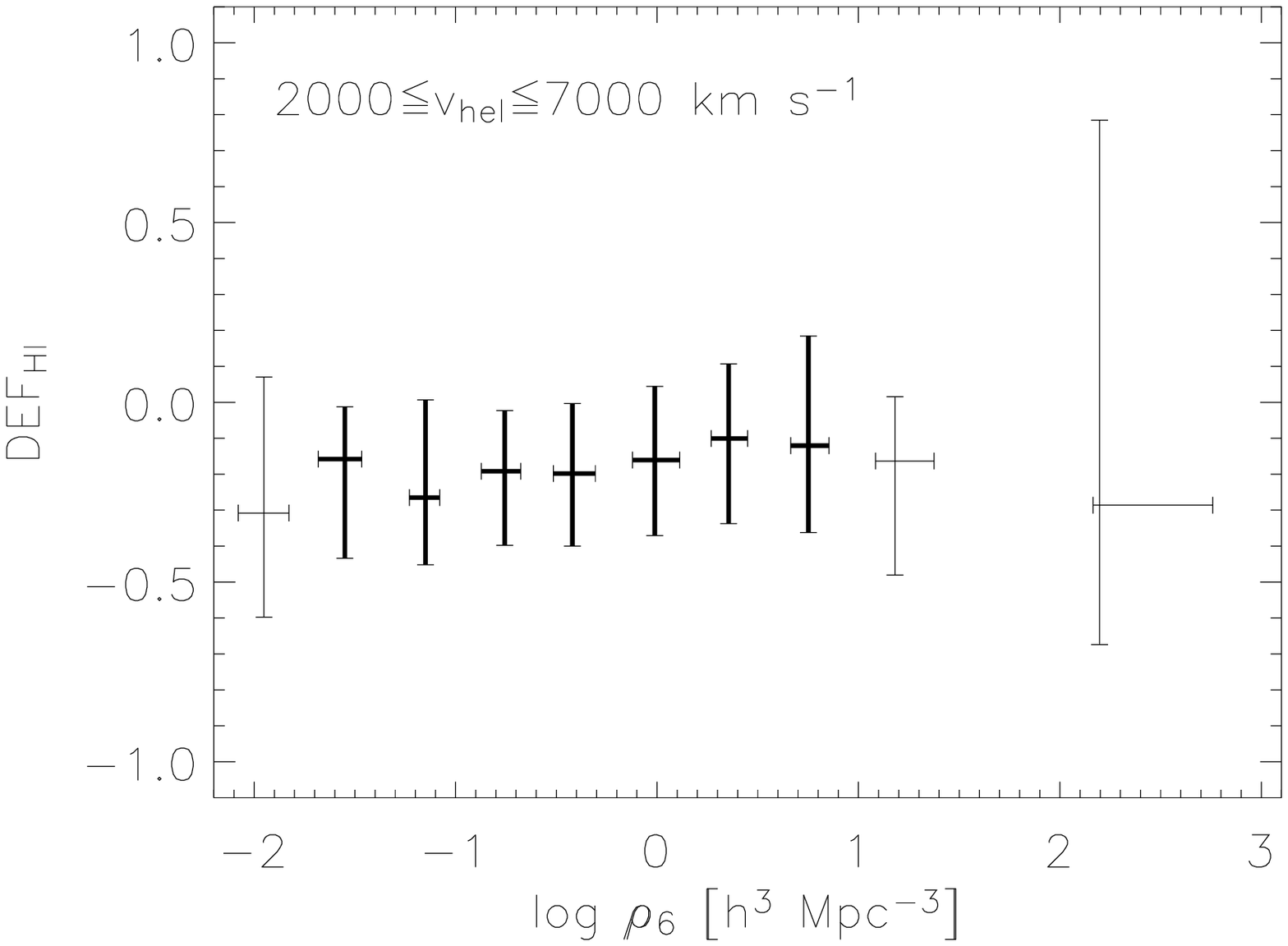}\\
\vskip0.15in
\plottwo{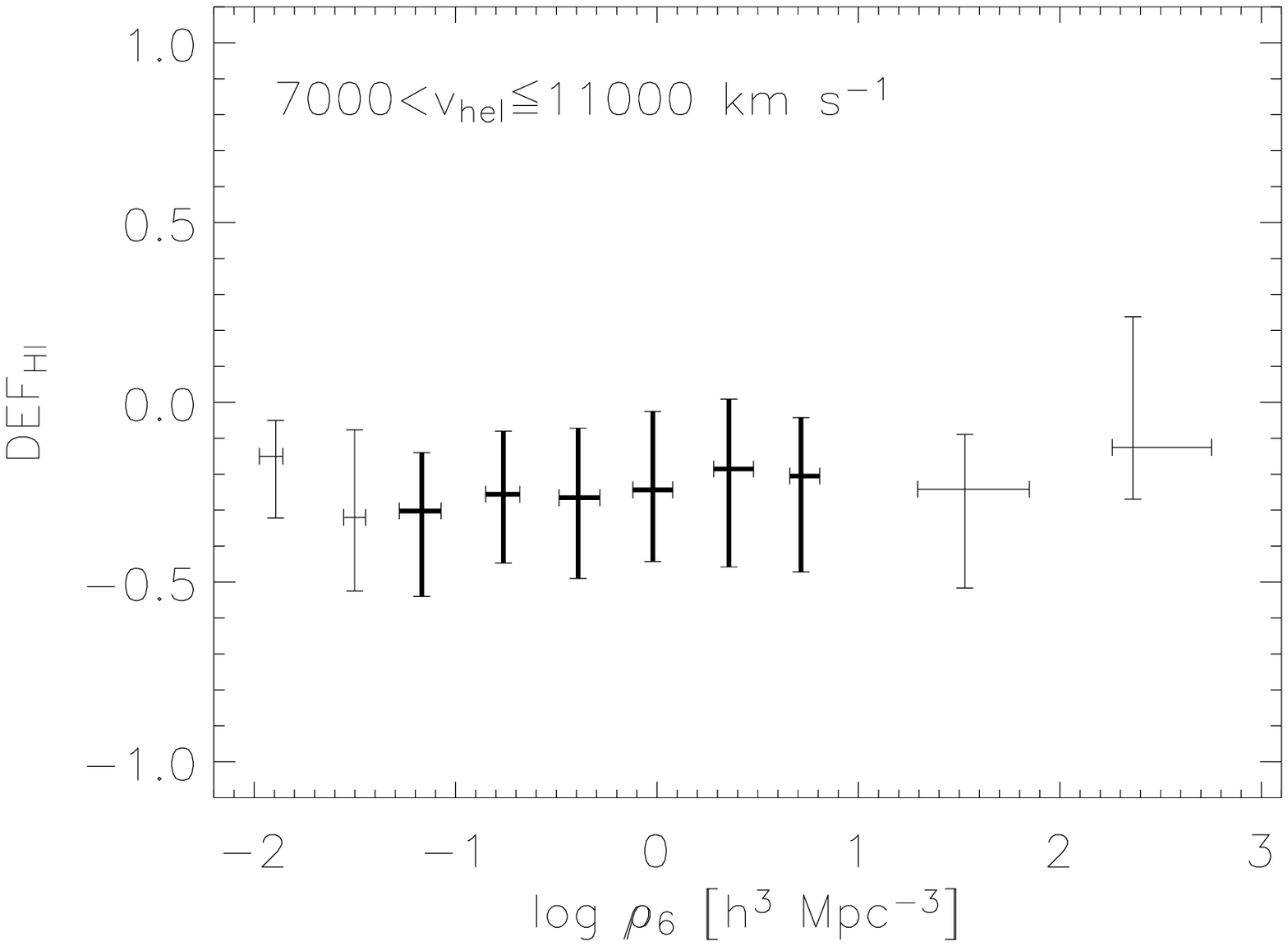}{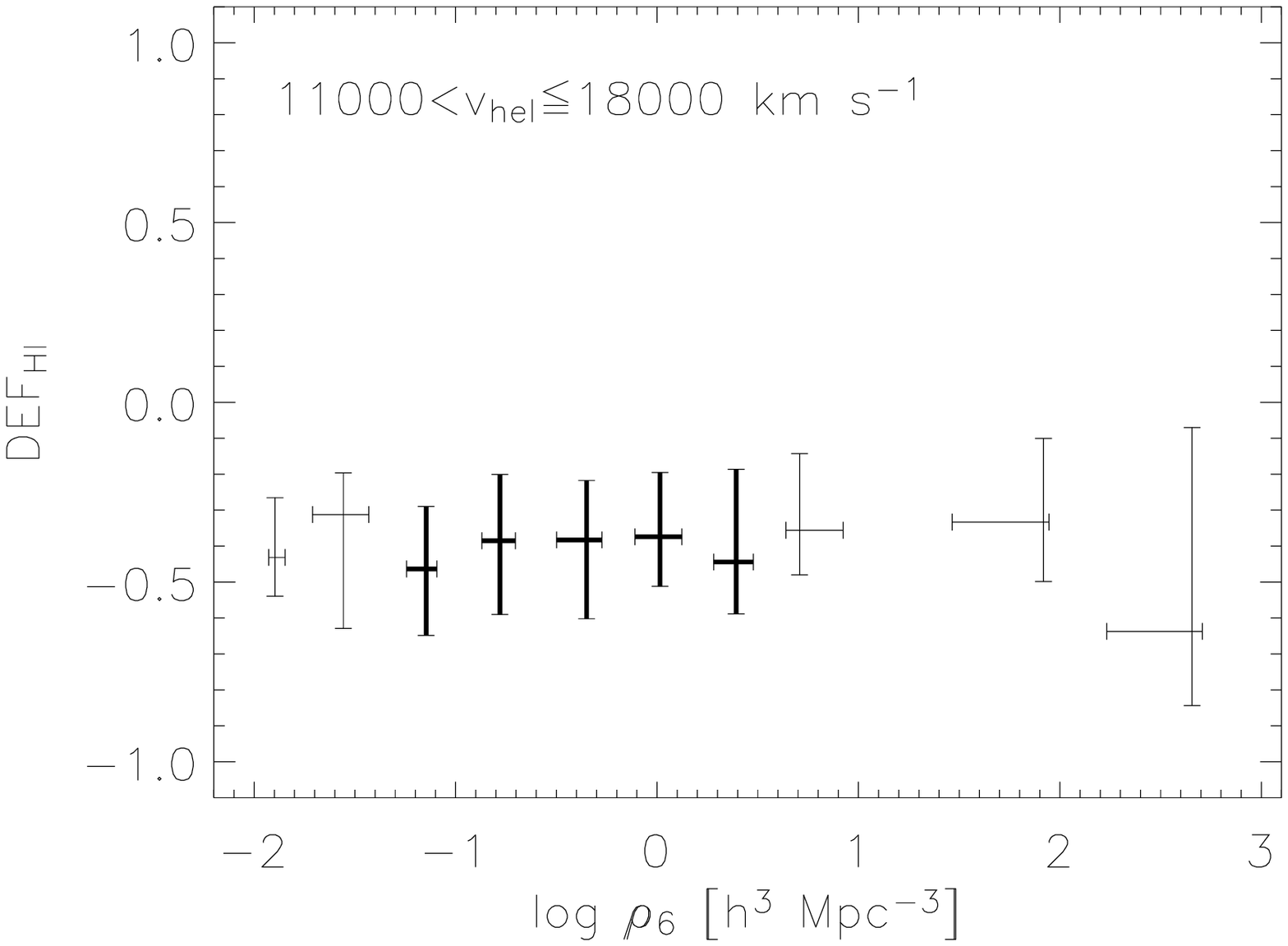} 
\caption{\small Medians and inter-quartile ranges of the
  \hi-deficiency parameter, $\df$, calculated using the standards of
  \hi\ content given by \citet{SGH96}, in equal bins of $\log \rho_6$
  up to $\log \rho_6 [h^3 \mbox{Mpc}^{-3}]=1$ . Above the latter value, only two larger equal
  bins are considered. Results for bins containing less than 25
  objects are represented by thinner lines. The different panels are
  for \alf\ detections of Sa--Sd type with radial heliocentric
  velocities between 2000 and $18,000$ \kms\ (top-left), as well as
  for the three subintervals: 2000--7000 \kms\ (top-right),
  7000--$11,000$ \kms\ (bottom-left), and $11,000$--$18,000$ \kms\
  (bottom-right).}\label{fig_def_vs_rho}
\end{figure}

We present in Figure~\ref{fig_def_vs_rho} a plot of the median $\df$
in bins of local number density. We show results inferred using
sources with radial velocities between 2000 and 18,000 \kms, as well
as separately for three sub-intervals of this velocity range. Looking
at the different panels one can see moderate oscillations in the
central values of $\df$. Nonetheless, these are always well within the
quoted uncertainties and therefore not statistically significant,
hinting at best a mild increase of gas deficiency with density. The
weak dependence of the gas content of \alf\ sources with the local
density is the consequence of limiting the effective integration time
of the survey to about 48 seconds per beam area. While \alf\ has been
designed to detect from massive spirals at $z\sim 0.06\;(\sim 250
\;\mbox{Mpc})$ to dwarfs with \hi\ masses as low as $\Mhi\sim
10^{7}\,\msun$ at the Virgo distance \citep[e.g.,][]{dSA07}, the
highly gas-deficient galaxies that are expected to reside in the
densest regions of the surveyed volume should escape detection unless
they are located nearby (see also Figure~\ref{fig_rhovsmu}). It is
also important to stress that the observed behavior of the gas content
with local density is not affected by the negative offset shown by the
inferred median $\df$ values. Such a bias obeys to several factors. On
one hand, as mentioned above, we are underestimating the expected
gaseous content of Scd--Sd galaxy types, which lowers the median $\df$
of our dataset about $0.15$ units. Moreover, it is the fact that \alf\
is detecting total 21-cm fluxes for individual galaxies that, on
average, are around $10\%$ larger than their AGC counterparts. This
reduces the average value of $\df$ by additional $0.05$ units. The
rest of the drop in the \hi\ deficiency can be attributed to
differences in the object selection with respect to
\citet{SGH96}. Note that the \alf\ survey extends further in redshift
than the sample used in this former work, which was centered on the
Pisces-Perseus Supercluster. This allows us to deal with a
significantly large number of high-\hi-mass objects. Besides, due to a
less robust selection of the control sample, we cannot discard the
presence in the \citeauthor{SGH96}'s data of a non-negligible
fraction of galaxies with a disturbed \hi\ mass, which might have
contributed to further reduce the standard values of this quantity
calculated in that paper.

\begin{figure}
\epsscale{0.95}
\plotone{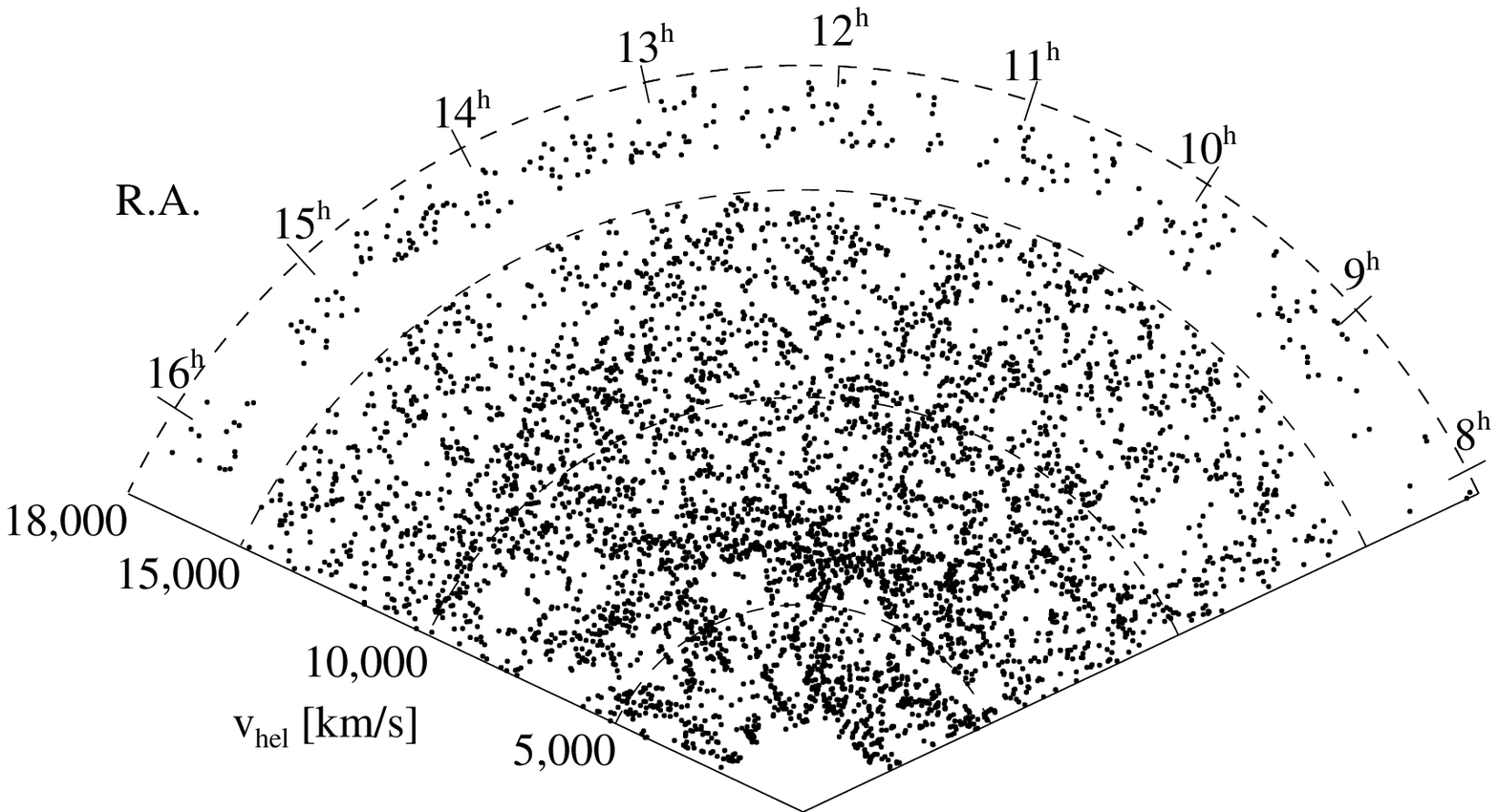}
\plotone{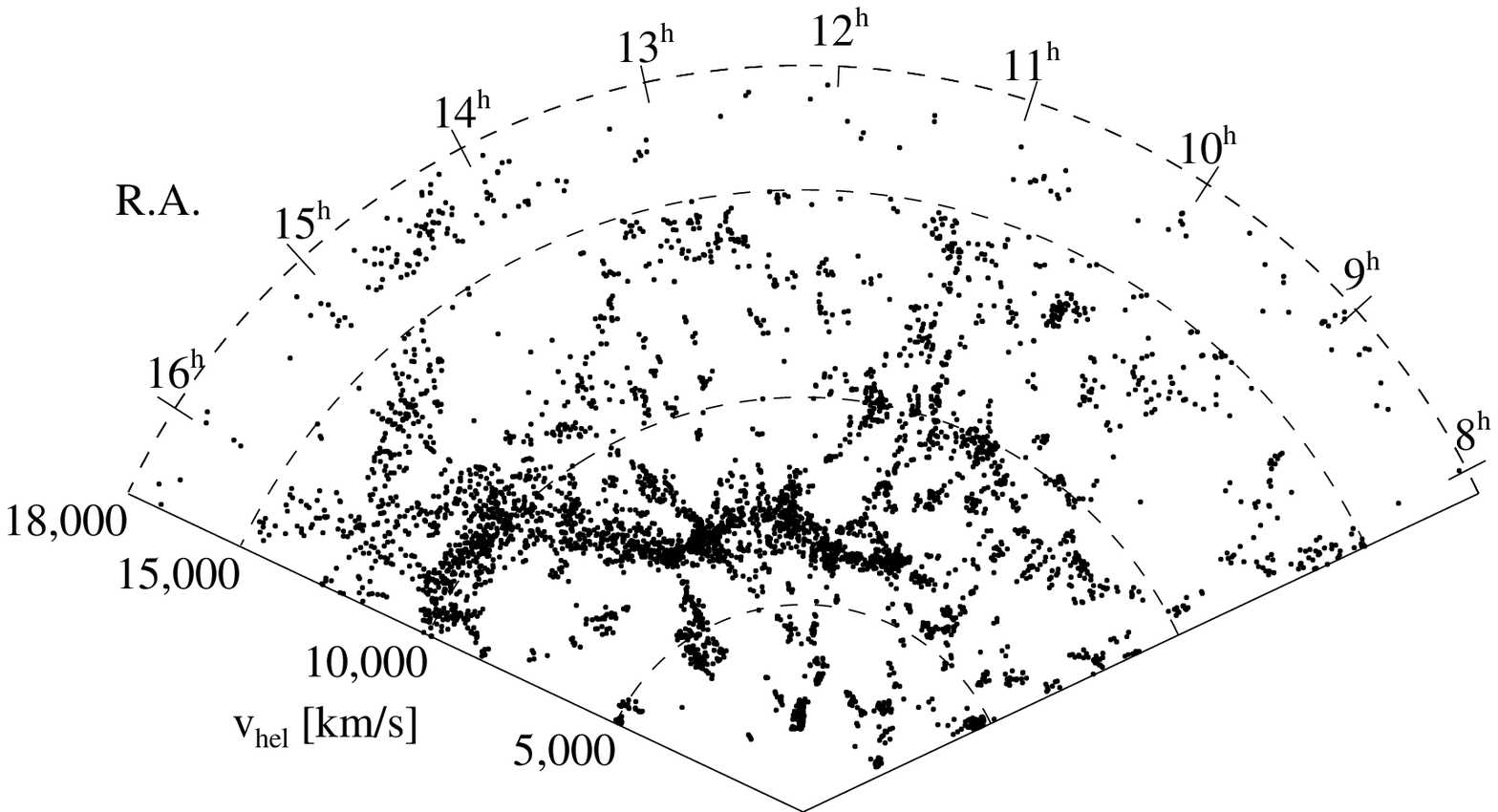}
\caption{\small Wedge diagrams in right ascension of the LDE sample of
  \alf\ galaxies defined in the text (top) and of those \alf\
  \hi\ detections lying in environments of higher density in the same sky region (bottom).}
\label{fig_wedge_ldr} 
\end{figure}

The lack of a clear \hi\ content-density correlation for the portion
of the \alf\ spring-sky survey under study notwithstanding, we prefer
to be conservative and, as indicated at the beginning of the section,
define a control sample that does not include galaxies located in the
vicinity of an overdensity. After testing different density
thresholds, we find that the condition $\rho_6\leq 0.5\,h^3$
galaxies~Mpc$^{-3}$ discards all SDSS galaxies with spectroscopic
information cataloged as cluster members in the surveyed region ---the
corresponding threshold based on the 2MASS galaxy distribution would
be $\mu_6\sim 10$~galaxies~deg$^{-2}$---, as well as more than 91 (96)
per cent of these objects with system-centric distances between
1.5--$3\,r_{\mathrm{vir}}$ (1.5--$2\,r_{\mathrm{vir}}$) and systemic
velocities within $1\,\sigma$ of the cluster mean, which may be
considered representative of the system outskirts population. This
threshold performs similarly well applied to \alf\ detections. The Low
Density Environment (hereafter LDE) \hi\ galaxy sample that results
from selecting those \alf\ detections obeying the above density
constraint has a total of 5647 objects. Wedge diagrams in right
ascension of the spatial distribution of this data set, as well as the
\hi\ detections discarded for lying in higher density environments,
are shown in Figure~\ref{fig_wedge_ldr}.

\subsection{Isolated Galaxy Samples}\label{isolated}

By selecting galaxies in low density regimes one may confidently
exclude objects associated with clusters and large groups, but cannot
guarantee that such galaxies have not interacted at some point with
nearby companions. That is the reason why we also studied
the possibility of defining control samples from the \alf\ catalog
that maximize the chances of dealing with galaxies whose
evolution has not been influenced by other objects in the recent
past. Here we build two samples of isolated galaxies by applying two
different kinds of isolation criteria.

As a first approach, we devised a selection technique for the SDSS
spectroscopic data set, similar to the one used by \citet{Pra03}, that
combines spectroscopic with photometric information to detect galaxies
without \emph{dynamically relevant} companions inside a given volume
in $z$-space. Thus, we classify a galaxy $i$ with apparent $r$-band
total magnitude $r_i$ at coordinates $(\alpha_i,\delta_i,z_i)$ as a
good candidate for a sample of isolated galaxies if it has no neighbor
$j$ with $r_j \le r_i+2.5$ mag within the cylindrical volume, centered
on its position, of radius and semi-axis
\begin{equation}\label{IG1}
R=R_0\psi^{-1/3}\ \ \ \ \ \mathrm{and}\ \ \ \ \ V=V_0\psi^{-1/3}+\sigma_{12}\;,
\end{equation}
respectively, where $R_0=0.7$~Mpc is the linking-length at the adopted
fiducial redshift of 2000 \kms, $V_0=H_0R_0$, and
$\psi(\alpha_i,\delta_i,z_i)$ is the selection function defined in
equation~(\ref{sel_func}) that keeps constant the probability of
finding a galaxy in a given volume regardless of distance. In
equation~(\ref{IG1}), the term $\sigma_{12}=350$ \kms\ is added in
order to account for the distortion on the intergalactic distances
along the line-of-sight induced by the one-dimensional pairwise
velocity dispersion of field galaxies \citep{Guz97, Lan02}. The
preference of a $0.7$~Mpc linking-length has been dictated from the
fact that it would take almost 3 Gyr for a galaxy with a peculiar
velocity of $175$ \kms\ to cover such a distance, while the magnitude
constraint is set to take into account neighbors that have at least
one-tenth of the target mass (assuming light is a proxy for
mass). This technique identifies 461 SDSS-spec galaxies that are also
\hi\ detections as members of this first isolated galaxy (hereafter
IG1) sample.

A second isolated galaxy (IG2) sample has been defined by applying the
sort of prescription typically used for photometric samples ---usually
much deeper than the spectroscopic ones---, which takes into account
in a semi-empirical way the distance to companions. Perhaps the best
known example is the nearest neighbor algorithm by \citet{Kar73} based
on the comparison of apparent angular diameters between candidate
isolated galaxies and their neighbors. Recently, \citet{All05} have
implemented a variation of \citeauthor{Kar73}'s technique to detect
isolated galaxies in the SDSS DR1 that we adopt here with a slight
modification. Thus, we consider a target galaxy $i$ with a $g$-band
total magnitude $g_i$ and $g$-band Petrosian radius $R_i$ to be
isolated if the projected sky separation $\theta_{i,j}$ with respect
to any neighboring galaxy $j$ satisfies the conditions
\begin{equation}\label{IG2}
\theta_{i,j} \geq 40 R_j\ \ \ \ \ \mathrm{and}\ \ \ \ \ \left| g_i-g_j \right|
\leq 2.5\;,
\end{equation}
where the 2.5 magnitude difference corresponds to about a factor 9 in
brightness for a galaxy with a flat surface brightness profile.  We
also took into account the fact outlined by \citet{All05} that
the application of these selection criteria to SDSS data effectively
reduces the magnitude range of the candidate isolated galaxies at both
ends (see, for instance, their Figures~11 and 12). Thus, by
considering only candidates with Galactic-extinction-corrected
$g$-band magnitudes in the range $15.5 \leq g_i \leq 19.5$, we 
extracted a total of 236 isolated \hi\ emitters\footnote{Had we
  followed the same approach, but imposing instead the more
  restrictive conditions $\left| g_i-g_j \right| \leq 3.0$ mag with
  $16.0 \leq g_i \leq 19.0$ mag used in \citeauthor{All05}'s paper,
  the number of retrieved isolated \hi\ galaxies would had drop to 67
  within the region surveyed by \alf.}.

Given that the spectro-photometric algorithm imposes the restriction
$r_i\leq 15.27$ on the magnitude of IG1 members, there is only a small
overlap in the range of apparent brightnesses of the galaxies that are
included in both subsets of isolated objects. In practice, this means
that only 7 \alf\ detections end up obeying simultaneously the two
isolation criteria. If we eliminate from the photometric prescription
the constraint $g_i\ge 15.5$ mag to make the overlap of both subsets
of isolated galaxies easier, then the number of IG2 members rises to
470. Even so, this enlarged sample includes only 83 of the 461
isolated galaxies selected by the spectro-photometric
algorithm. Therefore, isolated galaxies in $z$-space are not
necessarily considered so when their environment is examined in
projection and vice versa.

\begin{figure}
\epsscale{0.9}
\plotone{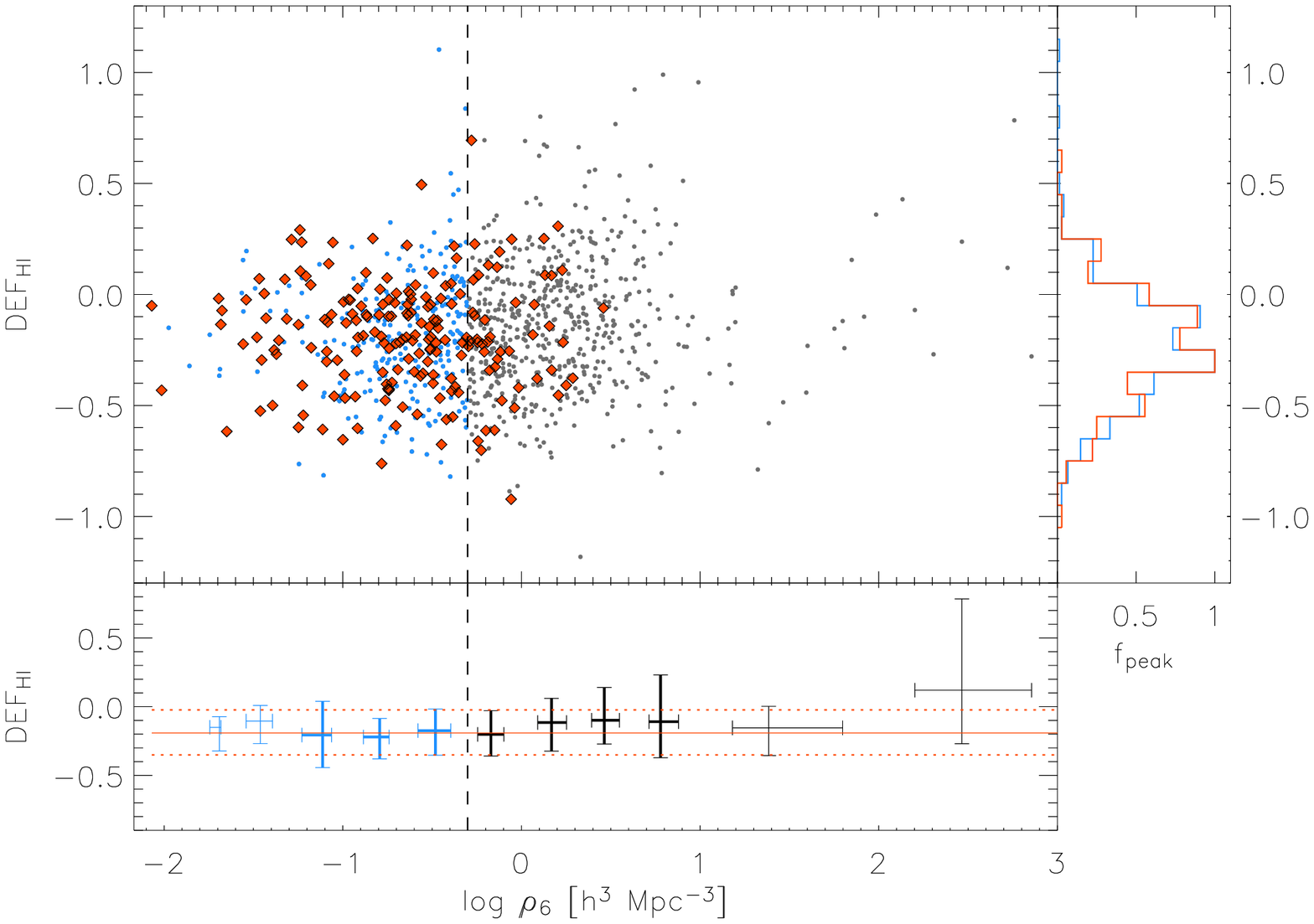}
\vskip0.45in
\caption{\small Top: $\df$, calculated using the standards of \hi\
  content given by \citet{SGH96}, vs.\ $\log \rho_6$. \alf\ detections
  of Sa--Sd type selected by a spectro-photometric isolation criterion
 (IG1 subset; see text) are shown as big red diamonds. Dots
  represent all Sa--Sd \alf\ detections spanning the same range of 
  $g$-band magnitudes than the isolated objects ($15.5\leq g_i \leq 
  19.5$). The subset of galaxies to the left of the vertical dashed 
  line (LDE galaxies) is highlighted in blue color. Bottom: 
  Corresponding medians and inter-quartile ranges of $\df$. Results 
  for all the \alf\ galaxies depicted in the main panel are 
  represented by error bars in equal bins of $\log \rho_6$ up to $\log 
  \rho_6[h^3 \mbox{Mpc}^{-3}]=1$, while above the latter value only 
  two larger equal intervals are considered. Results for bins
  containing less than 25 objects are drawn using thinner lines. The
  red horizontal lines show the median (solid) and upper and lower
  quartiles (dotted) of $\df$ inferred from all the galaxies in the
  IG1 subset. Right: Histograms normalized to peak values showing the
  distribution of $\df$ for the subsets of LDE galaxies (blue) and
  isolated sources (red). Similar results are obtained for the IG2
  subset.}\label{fig_def_vs_rho_isolate}
\end{figure}

As done in the previous section for the LDE \hi\ galaxy sample, we
also studied the environment of the \alf\ galaxies in the
IG1 and IG2 subsets. As an example, we display in
Figure~\ref{fig_def_vs_rho_isolate} a graphical representation of the
\hi-deficiency versus local density for IG1 galaxies. As before, we
focus only on sources having optical diameters and spiral types from
Sa to Sd listed on the AGC. As shown in this figure, while isolated
galaxies obtained through automated systematic searches avoid the
densest regions of the universe, a non-negligible fraction of them
inhabit regions of moderate local galaxy density having values of
$\rho_6$ larger than the upper density threshold adopted to define our
LDE \hi\ galaxy sample. This demonstrates that the definitions of
isolation and low density environment are not necessarily equivalent;
local number density measurements associated with isolated galaxies
can span quite a broad range of values \citep[for a similar
conclusion, see][]{Ver07}. Alongside this fact,
Figure~\ref{fig_def_vs_rho_isolate} also shows that although the
different approaches adopted to identify field galaxies assemble
samples of significantly different sizes, the corresponding gas
content distributions are quite similar. There is therefore no
evidence that our searches of isolated galaxies selected objects
that are particularly gas-rich compared to those in the LDE galaxy
sample.

\section{UNDERLYING PROPERTIES OF GALAXIES IN THE \hi\ CONTROL SAMPLES}\label{optical_properties}

Here we use several independent classification schemes based on the
photometric data for the galaxies in the three candidate
control samples defined in the previous section to provide more details
and increase our understanding of the objects that inhabit quiet
environments. These measures are all sensitive to the morphology and they 
are objective, easily reproducible, and applicable to data sets in
which the morphological classification by direct inspection of the
galaxy images is impractical, given the large number of objects
involved.

\subsection{Structure, Colors, and Morphology of the LDE
  Galaxies}\label{properties_LDE}

Since we are focusing our attention on subsets of non-clustered,
gas-rich galaxies, predictably of late type, we first apply the
criterion used by \citeauthor*{Mal09}~(\citeyear{Mal09}; hereafter
\citeauthor{Mal09}) to select disk galaxies from the SDSS. As stated
by those authors, almost pure disk samples can be obtained by
selecting galaxies that obey \emph{at least one} of these
requirements: S\'ersic index $n\leq 3$ or observed axial ratio
$b/a\leq 0.55$, where the parameter $n$ ---which in our case has
been extracted from the NYU Value-Added Galaxy Catalog
\citep[][]{Bla05}--- measures the shape of the observed $r$-band
luminosity profile of a galaxy fitted using the S\'ersic $R^{1/n}$
formula with elliptical isophotes. We note that these constraints
include S0 galaxies in the 'disk' class, because the presence of a
disk makes their measured axial ratio more important in determining
their apparent inclination than the intrinsic ellipticity of the
spheroidal component. According to \citeauthor{Mal09}, the
completeness of their classification scheme reaches $70\%$ for disk
galaxies, with a reliability of $\sim 95\%$ when applied to SDSS
data. The completeness is the fraction of galaxies of a given type
that are successfully selected from the original sample by the
classification scheme. The reliability is the fraction of galaxies of
the desired type in the selected subset.

In good agreement with expectations, Table~\ref{tab_maller} shows that
the fraction of members of the SDSS-spec sample in low density
environments ($\rho_6\leq 0.5\,h^3$ galaxies~Mpc$^{-3}$) that obey
\citeauthor{Mal09}'s criterion is very high ($\sim 85\%$ on average),
and even higher for the subset of \alf\ detections ($\sim 92\%$). Our
global percentages of late-type objects are, however, slightly lower
than those obtained by \citeauthor{Mal09}. This result can be
attributed to the fact that they were inferred for a very shallow data
set with $M_r<-20.6$ mag (note that the agreement between our results
and \citeauthor{Mal09}'s becomes almost perfect if we restrict the
comparison to our lowest velocity bin). In principle, the decrement in
the fraction of galaxies assigned to the disk class with increasing
radial velocity illustrated in Table~\ref{tab_maller} can be
interpreted as an indication of the fact that most luminous galaxies
tend to be of early type, in qualitative accordance with previous work
\citep[e.g.,][]{Str01,Bla03,Bal04}. Yet we find that in the highest
velocity bin $\sim 10\%$ of the \alf\ galaxies are not classified as
disks, while at these relatively large distances \hi\ detections
should arise almost exclusively from late-type objects: at $11,000$
\kms\ the minimum \hi\ mass detectable in the survey is $\log (M_{\rm
  {H\,\scriptstyle I},min})\sim 9.3\,\msun$. Accordingly, our results
seemingly highlight the difficulties in morphology assignment that
selection rules like \citeauthor{Mal09}'s, based on global measures of
galaxy structure, encounter for faint objects due to the inherent
noisiness of the images.

The inverse light concentration index, $\ici=R50/R90$, is another
measure of galaxy structure that can be used as a morphological
divider \citep{Shi01,Str01}. As with the S\'ersic index, this
parameter is expected to be correlated with galactic
morphology. Indeed, the two indexes can be interchanged for an ideal
galaxy with a S\'ersic profile in the absence of seeing. Early-type
galaxies have a dominant bulge component, so their light is
concentrated, while late-type galaxies, which harbour bulges of
varying size, have their light more dispersed due to the presence of
an extended exponential disk component and hence larger $\ici$
values. In practice, however, the concentration index is a
morphological separator that is useful for broadly dividing galaxies
into two populations: early-type (E, S0, Sa) and late-type (Sb, Sc,
Irr) objects. We follow \citet{Str01} and adopt an $\rici=0.38$
separator in the $r$-band, which provides a cut giving equally
complete subsamples of both morphological populations, while at the
same time it reduces the contamination of the late-type galaxy class
by early-type objects to $\sim 12\%$. When applied to galaxies in
LDEs, we find that most galaxies have $\rici>0.38$, indicating the
dominance of disks. As was to be expected, this tendency is again
more marked for \alf\ detections, about $82\%$ of them fall into the
light-concentration-index late-type class, than for the optical
subset, which has a late-type galaxy fraction of $\sim 70\%$.

Given the correlation between color and morphology, early-type
galaxies being generally redder than late-types, galaxy colors can
also be used as a morphological classification tool, as well as to
provide a coarse separation of galaxies in terms of their star
formation activity. We adopt the classification scheme of
\citet{Bal04}, which uses the SDSS model magnitudes \citep{Sto02} to
measure color from the difference between the fluxes in the $u$ and
$r$ filters and to calculate the absolute magnitude in the $r$ band,
$M_r$, for which we adopt the CMB distance inferred in
Section~\ref{distances}. As well as being an optimal color for
separating galaxies into red and blue classes \citep{Str01}, $(u-r)$
is a good indicator of star formation activity \citep{DZ09}, while the
$r$ filter contains the peak of the light curve for most SDSS
galaxies. In this analysis, we do not apply $K$-corrections since the
samples studied are shallow.

\begin{figure}
\epsscale{1}
\plotone{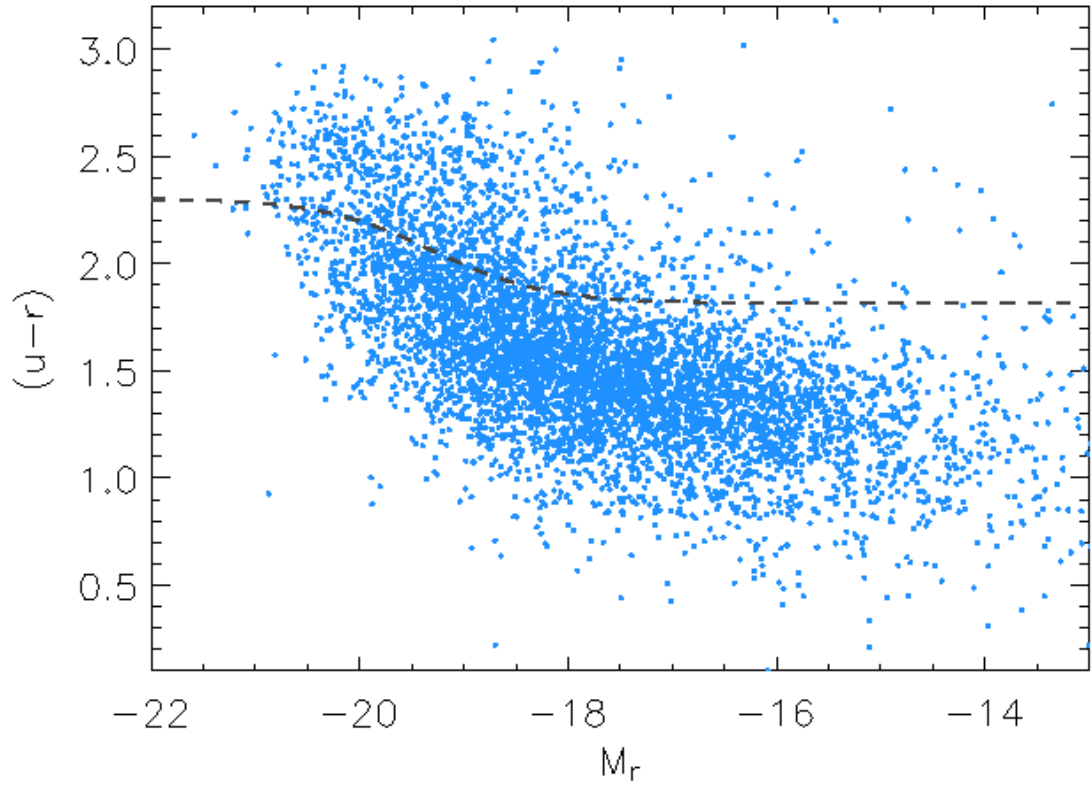}
\caption{\small $(u-r)$ color versus absolute magnitude in the
  $r$-band for \alf\ galaxies in the LDE subset. The dashed curve
  represents the red-blue population separator adopted by
  \citet{Bal04}. Colors are uncorrected for internal
  extinction.}\label{fig_color-magnitude}
\end{figure}

Figure~\ref{fig_color-magnitude} shows the color-magnitude
diagram for \alf\ detections in the LDE sample with the
\citeauthor{Bal04} delimiter,
\begin{equation}
(u-r)=2.06-0.244\,\mbox{tanh}\left(\frac{M_r + 20.07}{1.09}\right)\;,
\end{equation}
superposed. Not surprisingly, we see quite a few more sources on the
blue, active population side of the divider (below the curve) than on
the red, passive side: $83\%$ and $17\%$ of the total,
respectively. Moreover, we note that most of the \hi\ galaxies above
the separator reside in an area of the color-magnitude diagram around
$M_r\lesssim -20$ mag and $(u-r)\gtrsim 2.3$ mag where the blue and
red galaxy populations identified by \citet{Bal04} overlap in this
space (see their Figure 9). Therefore, it is not unlikely that many of these galaxies are actually blue interlopers. Indeed, if we repeat the same
exercise after correcting the SDSS magnitudes and colors to face-on
values, following respectively \citet{Sha07} and \citet{Mas10a}, and
keeping the same separator, the fraction of \hi\ detections assigned
to the blue group increases to $92\%$. This highlights the
importance of defining morphological separators corrected for internal
extinction, since the effects of dust on the colors and concentration
of late-type galaxies can be significant for edge-on systems.

A slightly coarser color division, but one that facilitates the
analysis of the data, is provided by the luminosity-independent
$(u-r)$ color cut of 2.22 mag ---also uncorrected for internal
extinction--- suggested by \citet{Str01}. In the four panels of
Figure~\ref{fig_color-LDE}, we show the observed color distributions
of the SDSS-spec galaxies in LDEs separated into \hi-detections and
non-detections for different bins of radial heliocentric velocity. As
illustrated by the histograms, the global fraction of SDSS galaxies in
LDEs that are of red-/early-type becomes notably higher with
increasing distance, in agreement with the well-known
color-luminosity-morphology(-star formation activity) correlation (see
also Figure~\ref{fig_color-magnitude}), which appears to hold
regardless of local galaxy density \citep*[e.g.,][]{BLB08}. Yet it is
obvious from these plots that the observed behavior arises almost
exclusively from the contribution of galaxies undetected in \hi. For
these systems, the fraction of red objects, according to the
\citet{Str01} separator, grows steadily from less than $15\%$ in the
lowest velocity bin, to about $39\%$ in the highest one, where $\sim
96\%$ of the SDSS galaxies redder than the divider are \hi\
non-detections. In sharp contrast, field \hi\ emitters consist of
(mostly blue) objects whose color distribution shifts only very
slighty redwards with increasing distance, to the point that the ratio
between the blue and red fractions for this subset remains nearly
constant at $\sim 11\pm 4\%$ in all redshift bins ($\sim 5\pm 2\%$ if
one uses colors corrected for internal extinction) and, hence,
approximately independent of the mean \hi\ mass of the
detections. This result contrasts with previous claims, based on
optically-selected sources, that blue exponential systems, just as
with red concentrated ones, become monotonically redder with
increasing luminosity \citep[e.g.][]{Bla03,BLB08} and suggests that
the color and \hi\ mass of gaseous galaxies cannot be very strongly
related (see also Section~\ref{properties_isolated}). In Paper II we
will return to this issue, which also has fundamental implications for
the theory of galaxy formation regarding the number of independent
physical parameters that determine the properties of individual
galaxies \citep{GPB96,Dis08,Gav09}.

The Galaxy Zoo 1 (GZ1) clean catalog \citep{Lin10}, which provides
visual morphological classifications for nearly 900,000 SDSS galaxies,
offers the possibility of directly inspecting the basic morphology of
a highly significant fraction of our \hi\ detections. To this
end, we focused our attention on the subset of 4599 GZ1
objects that are members of the SDSS DR7 spectroscopic survey and are 
also included in our LDE \hi\ galaxy sample. Following
\citeauthor{Lin10}, we were rather strict in assigning GZ1
morphologies to our \alf\ objects by requiring that they have a 'spiral'
or an 'elliptical' flag in the clean catalog, i.e., with more than $80\%$
confidence in category assignation after debiasing corrections. All other
galaxies were classified as of uncertain type ---as shown in
Figure~\ref{fig_gz-LDE}, most of the \alf\ detections with uncertain
GZ1 morphology are objects with $n<2$ but $b/a\gtrsim 0.4$. Comparison 
with the Hubble types listed in the AGC indicates that the reliability 
of this classification scheme for the spiral class is higher than  
$90\%$, with a $\sim 3\%$ contamination from S0/S0a galaxies, in good 
agreement with estimates by \citet{Lin08} and \citet{Bam09}. In this 
manner, we infer that $\sim 52\%$ of our \hi\ detections can 
confidently be assigned a spiral (disk with spiral arms) morphology, 
whereas only $\sim 1\%$ are galaxies with a high probability of 
belonging to the elliptical class (E and S0 galaxies). The fact that 
all the objects in the latter class have $b/a>0.5$ and that most of 
them have $n>3$ ($r$-band values) implies that prescriptions like that 
used by \citeauthor{Mal09} are very effective at separating out objects 
of early-type morphology. Indeed, we find that only 19 of the 4057
\citeauthor{Mal09} disks identified in this subset of GZ1 members are
classified by the Galaxy Zoo as bona fide ellipticals. We note in
addition that just 78 of the 4599 members of the LDE-GZ1 \hi\ subset
have a weighted-merger-vote fraction larger than $40\%$ \citep{Dar10}.

We extended this a step further by examining by eye the SDSS images of
those LDE gaseous galaxies that show optical early-type
characteristics according to both their color and S\'ersic index. We
concentrated our efforts on objects with extreme inclinations that
meet the following criteria: $(u-r)>2.22$ mag, $n>4.0$, and $b/a<0.5$
or $b/a>0.85$ (the latter two parameters measured in the $r$-band). In
agreement with \citeauthor{Mal09}, we found that most of our edge-on,
red, concentrated \hi\ sources have a clear disk-like morphology
suggesting that they owe their early-type classification to large
amounts of dust reddening. This morphology also appears to be the
dominant one among the face-on gas-rich galaxies identified with large
S\'ersic indexes and red colors, in which luminous red central regions
and large bars are plentiful. We speculate that these latter objects
represent a stage of spiral evolution that is not as advanced as that
assigned to the population of red spirals recently studied by
\citet{Mas10b}, since ours still have very blue, but faint, outer
disks and are quite rich in neutral gas (several of them have $\Mhi >
10^{10}\,\msun$), in good agreement with the findings of the GALEX
Arecibo SDSS Survey \citep{Wan10}. Therefore, we are left with only a
handful of cases that may correspond to genuine gas-rich early-type
galaxies (ETGs) in which the cold gas could have a recent external
origin. While the scarcity of this sort of 'hybrid system' in our LDE
sample is consistent with the very few large ETGs detected in \hi\ in
the Virgo cluster region \citep{dSA07}, it seems to be at odds with
the preliminary results of the ATLAS$^{\mathrm{3D}}$ survey of 262 nearby
galaxies that lack spiral arms in optical images \citep{Ser10}, which
hint at the possibility that isolated ETGs embedded in massive
\hi\ disks are relatively abundant.

The results discussed above, plus the fact that the automated
classification of galaxies into two major populations using extensive
stellar parameters leads to subdivisions that are partially
contaminated by members of the other class, and therefore to the
standardization of the trends, lead us to conclude that \hi\ emission
offers the most reliable (and clear) way to determine whether a galaxy
has genuinely an early-type or late-type morphology.

\subsection{Structure, Colors, and Morphology of the Members of the
  IG1 and IG2 Subsets}\label{properties_isolated}

The exercises reported in the previous section for \alf\ detections and
non-detections in LDEs were repeated for the samples of
isolated galaxies.

\begin{figure}
\epsscale{0.49}
\plotone{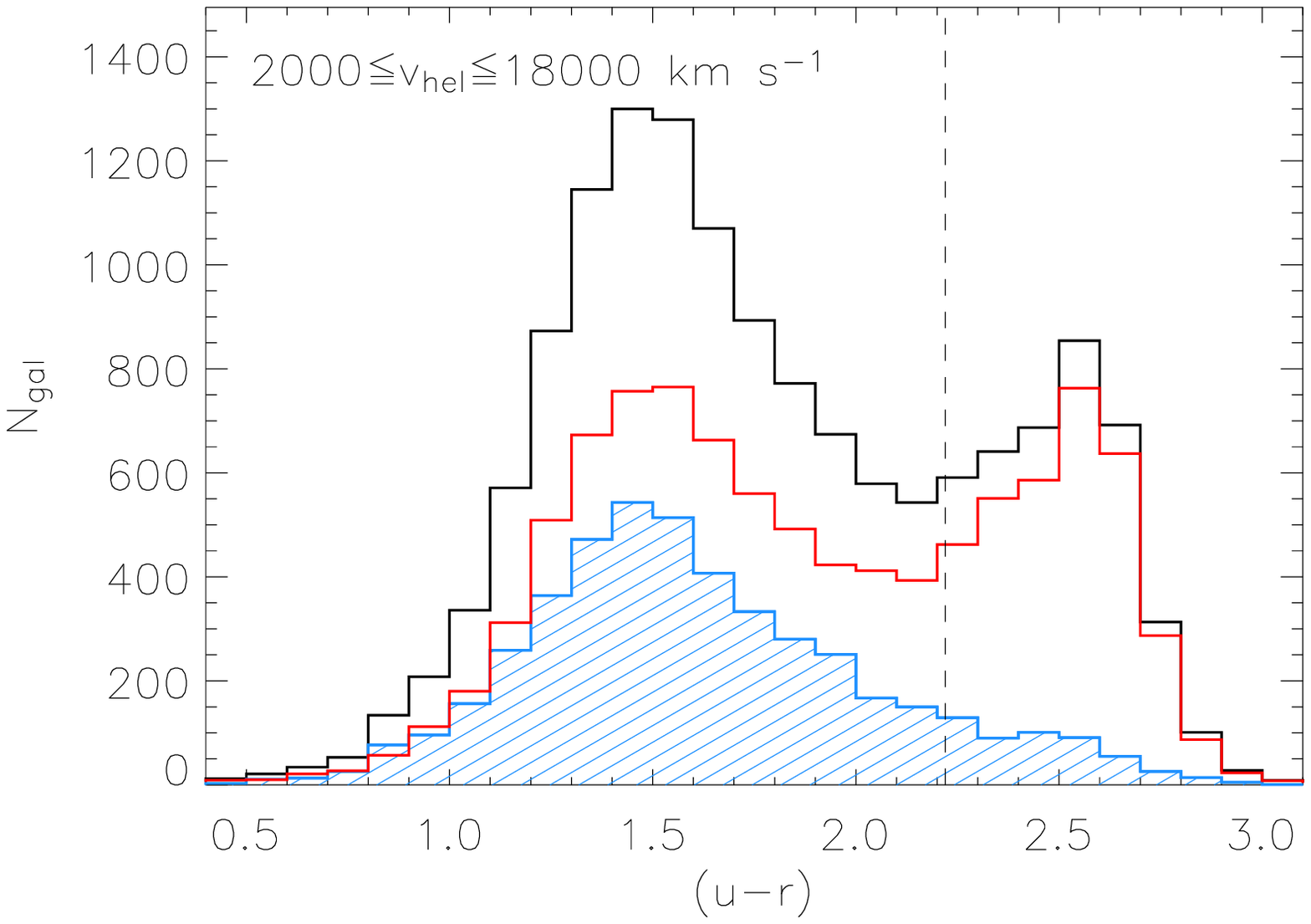}
\plotone{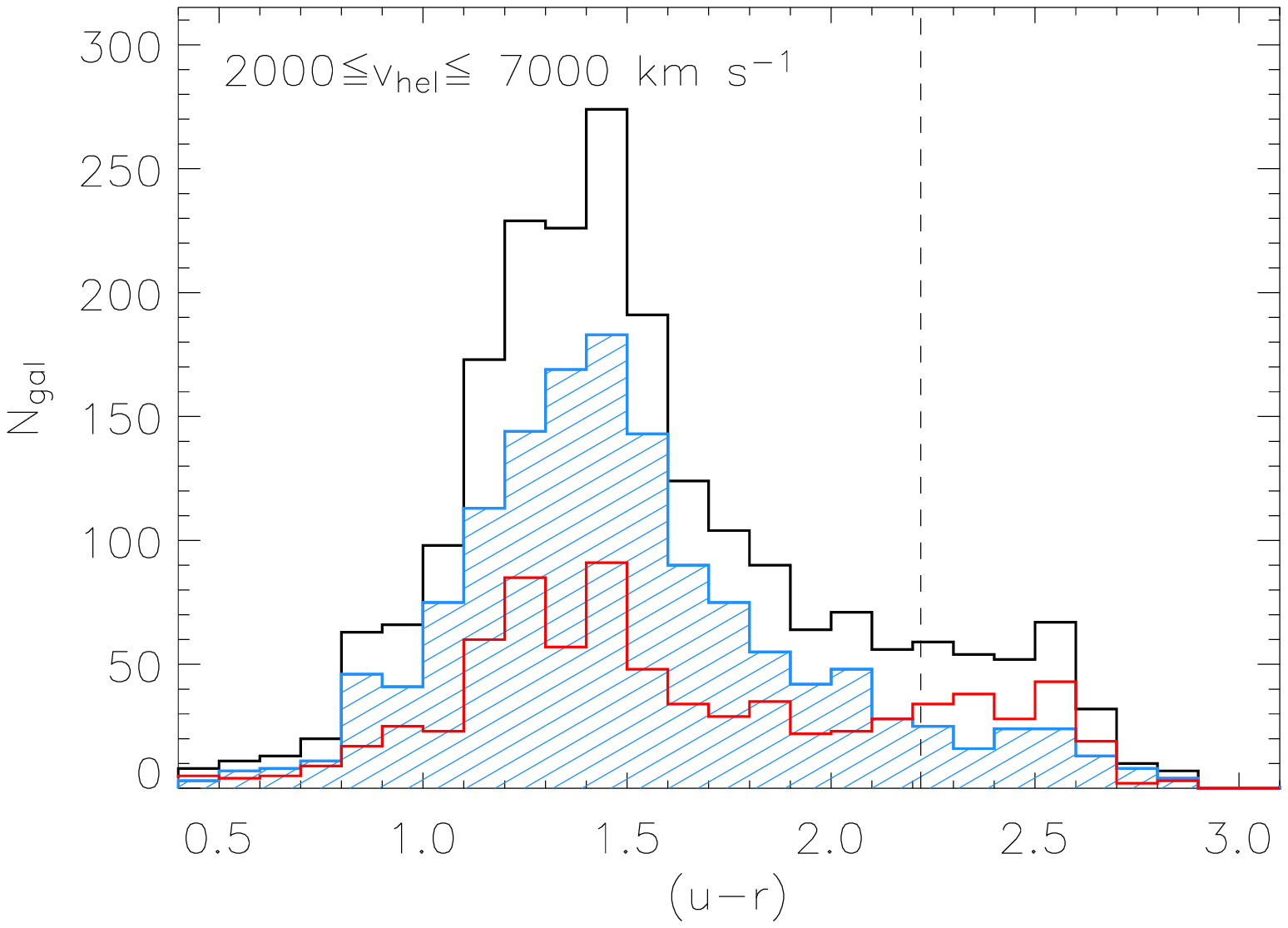}
\plotone{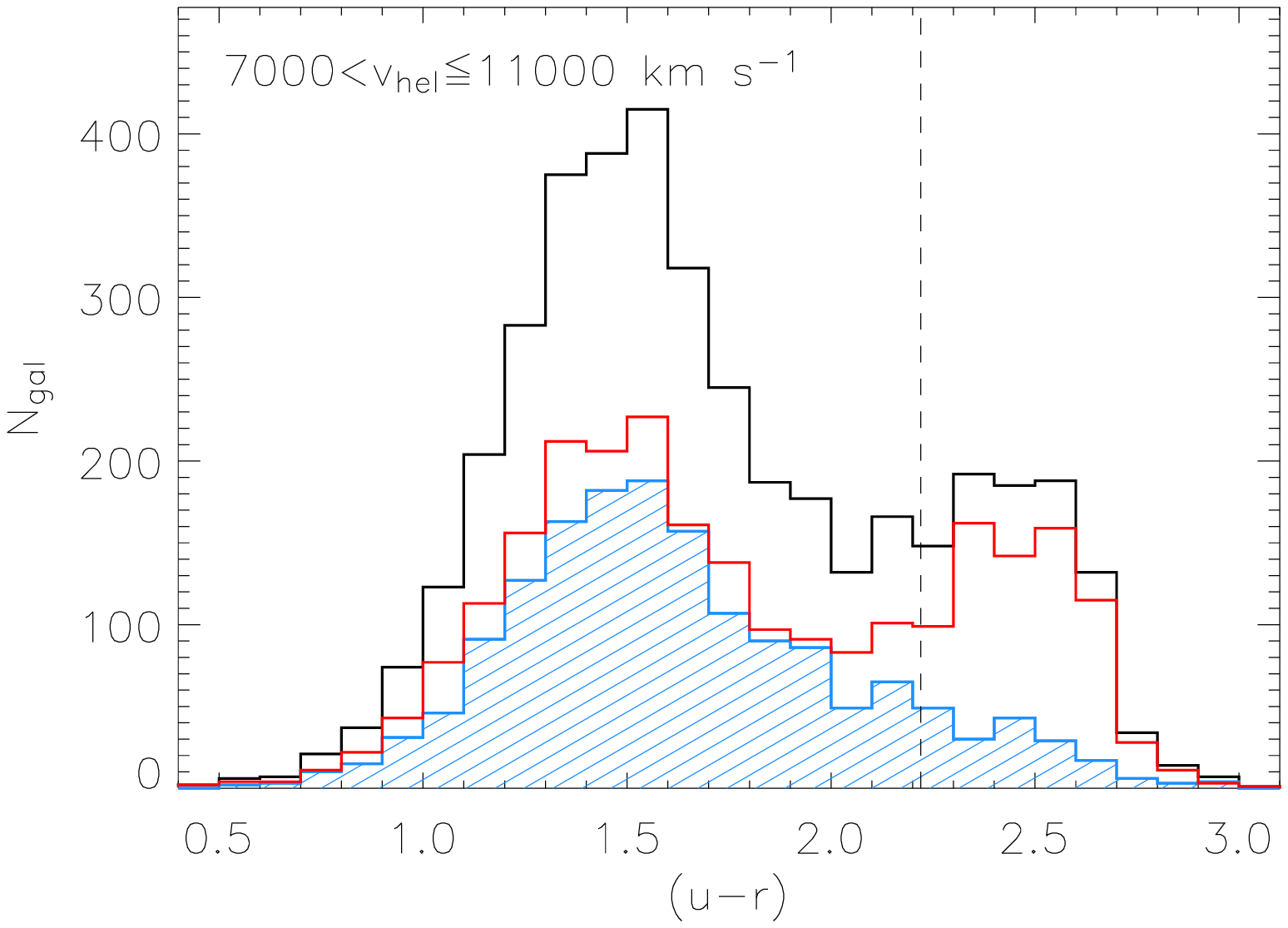}
\plotone{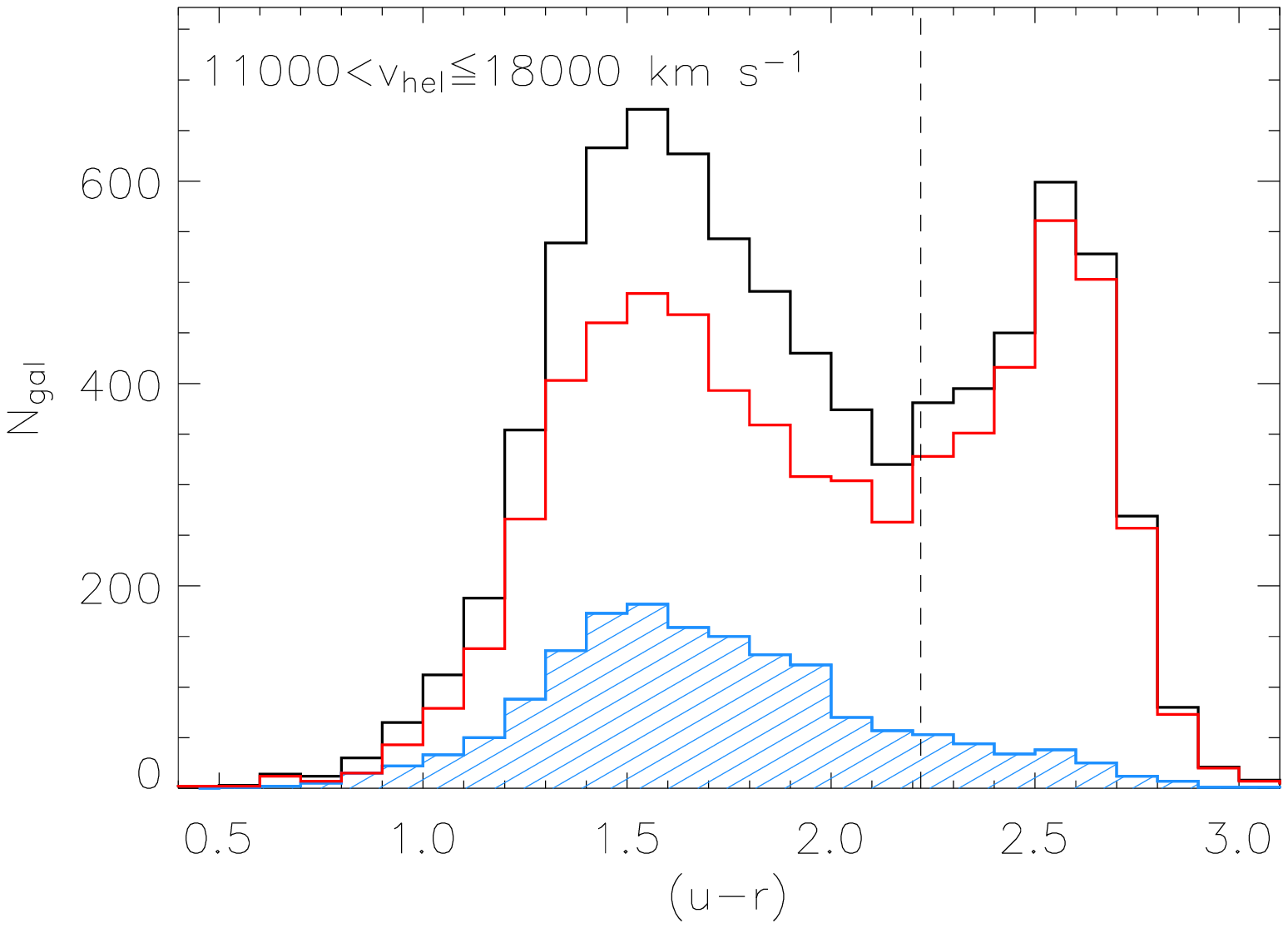}

\caption{\small $(u-r)$ color distribution for SDSS galaxies in LDEs
  for different bins of radial heliocentric velocity. The global
  histograms are split into \hi\ detections (dashed blue) and
  non-detections (red line). The vertical dashed line represents the
  red-blue population separator adopted by
  \citet{Str01}.}\label{fig_color-LDE}
\end{figure}

\begin{figure}
\epsscale{0.7}
\plotone{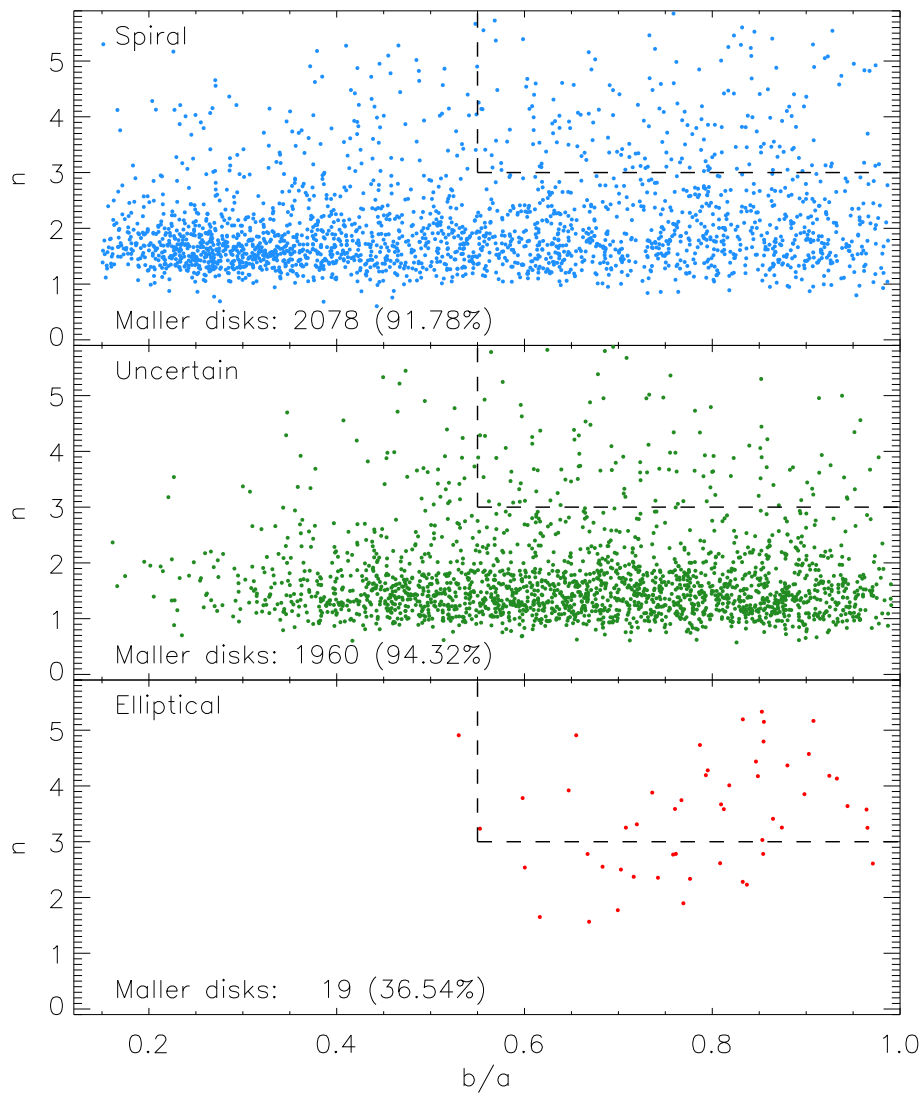}
\vskip0.45in
\caption{\small Distribution of GZ1 morphologies for \alf\ LDE
  detections in the S\'ersic index-observed axis ratio space
  ($r$-band): Spiral (top), Uncertain (middle) and Elliptical
  (bottom). In each panel, the top-right corner region delimited by
  dashed lines encompasses those objects not verifying
  \citeauthor{Mal09}'s disks criterion.}\label{fig_gz-LDE}
\end{figure}

\begin{figure}
\epsscale{0.7}
\plotone{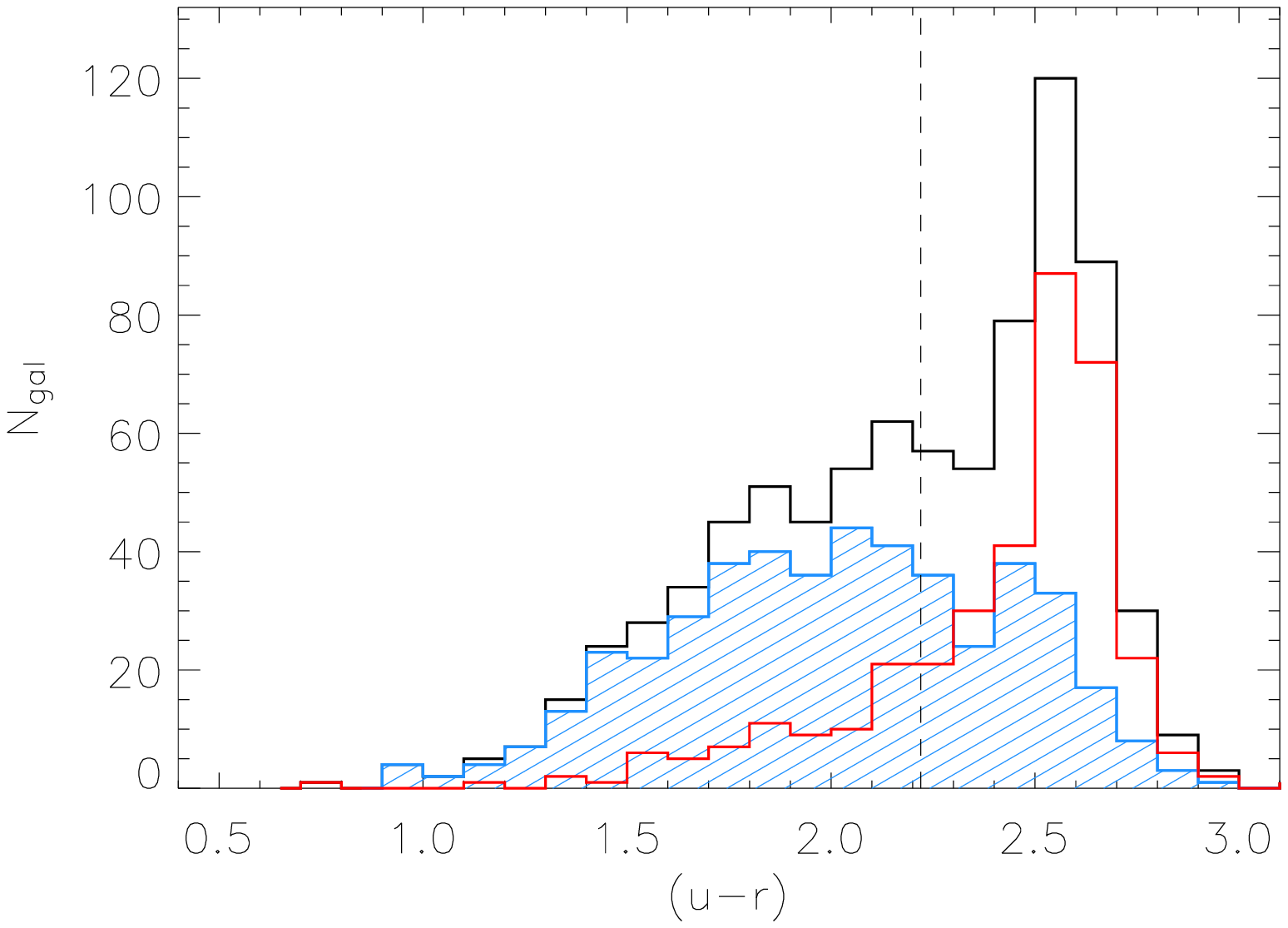}
\plotone{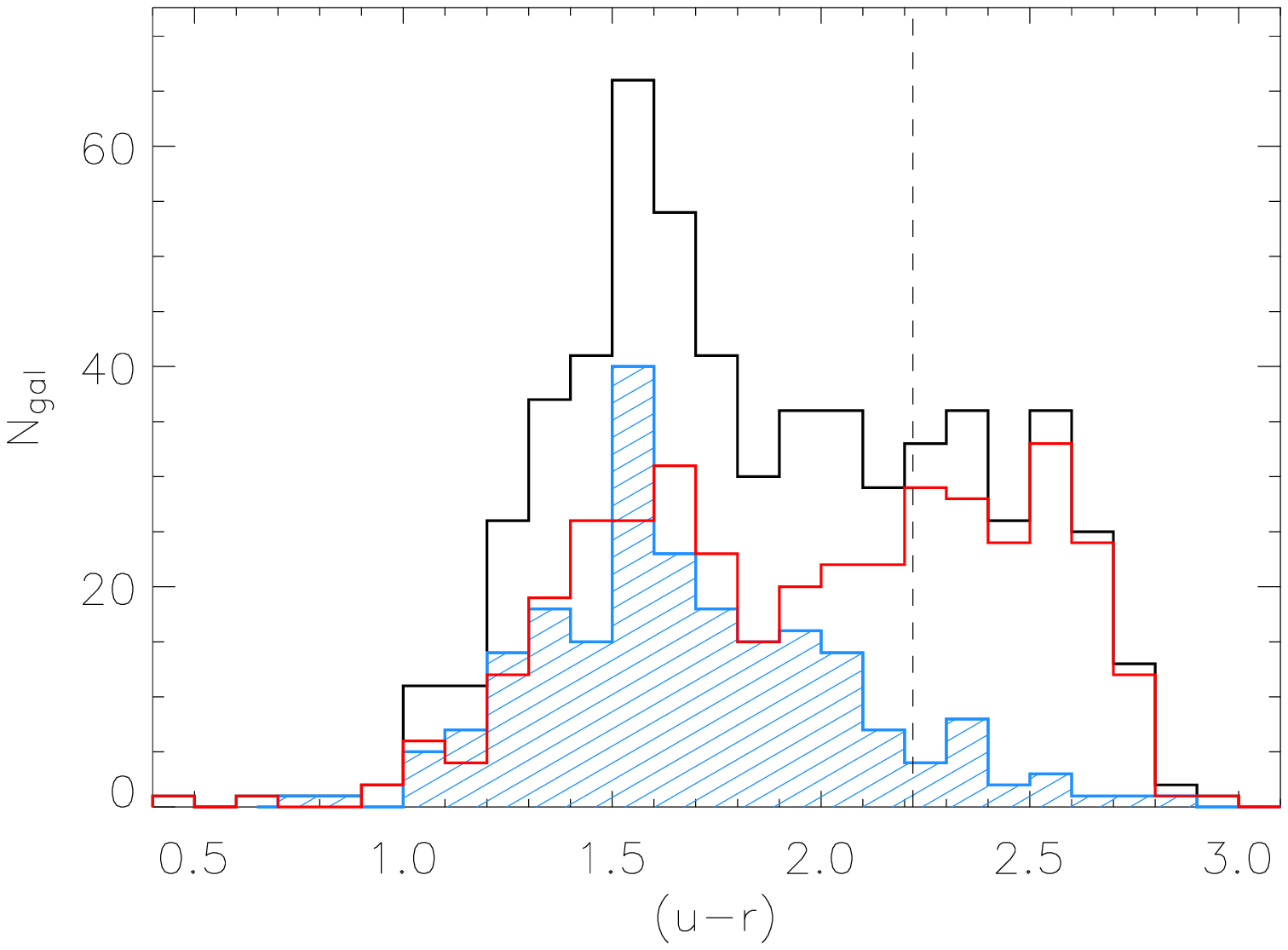}
\caption{\small $(u-r)$ color distribution for SDSS galaxies obeying
  the spectro-photometric (top) and photometric (bottom) isolation
  criteria (see text). The global histograms are split into \hi\
  detections (dashed blue) and non-detections (red). The vertical
  dashed line represents the red-blue population separator adopted by
  \citet{Str01}.}\label{fig_color-isolate}

\end{figure}

As illustrated in Figure~\ref{fig_color-isolate}, we find that,
regardless of the isolation criterion adopted, the population of \hi\
detections is notably bluer than that of non-detections, in good
agreement with our results for galaxies in LDEs. There are, however,
significant differences in the color distributions of both samples of
isolated galaxies. Thus, while the colors of the IG2 members mimic the
behavior shown by the LDE galaxies at the median depth of the \alf\
survey (bottom-left panel of Figure~\ref{fig_color-LDE}), the IG1
members display nearly unimodal distributions ---there is a hint of
bimodality in the histogram of \hi\ detections--- strongly skewed to
the red. Likewise, we find that IG1 galaxies tend to have both smaller
light concentration indexes and a lower fraction of
\citeauthor{Mal09} disks than galaxies in both the IG2 and LDE data
sets, as would be expected from a population more dominated by
early-type objects. Indeed, if we restrict our attention to just the
\alf\ galaxies with Hubble types listed in the AGC, we find that the
latter two samples are considerably richer in Scd and Sd spiral
subtypes than the IG1 subset is.

\begin{figure}

\begin{tabular}{cc}
 
\includegraphics[width=0.5\textwidth]{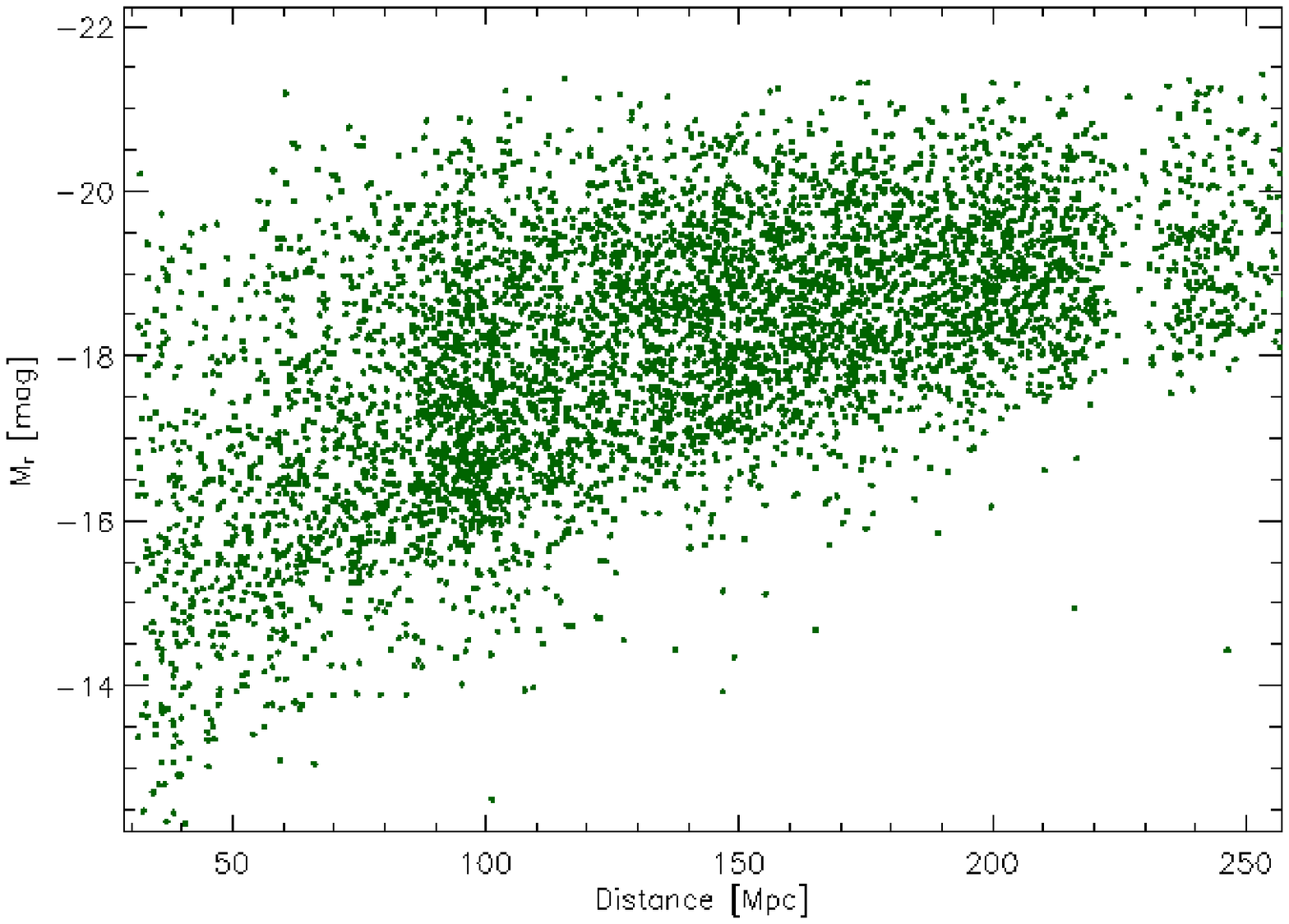} &
\includegraphics[width=0.5\textwidth]{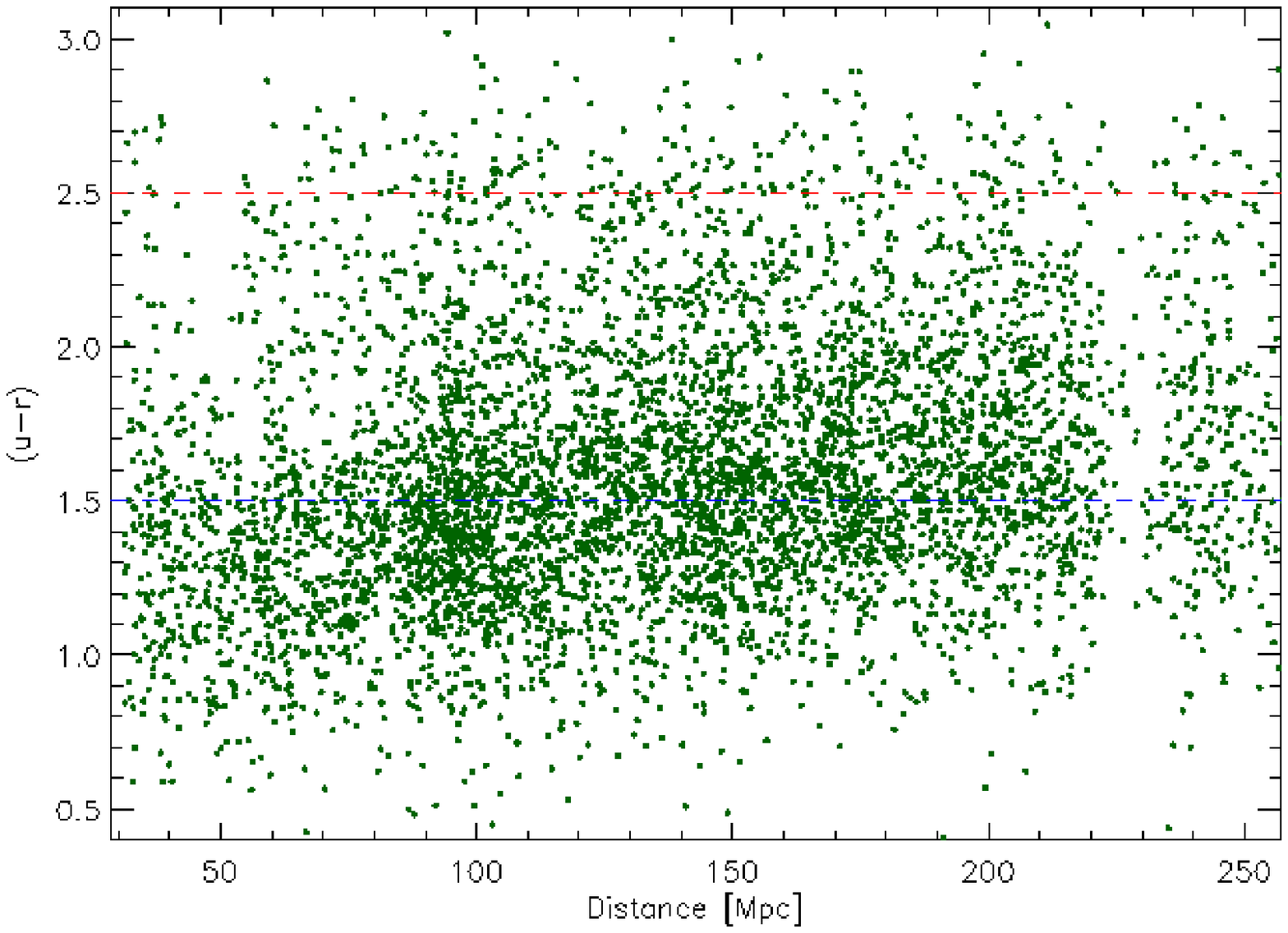}\\

\includegraphics[width=0.5\textwidth]{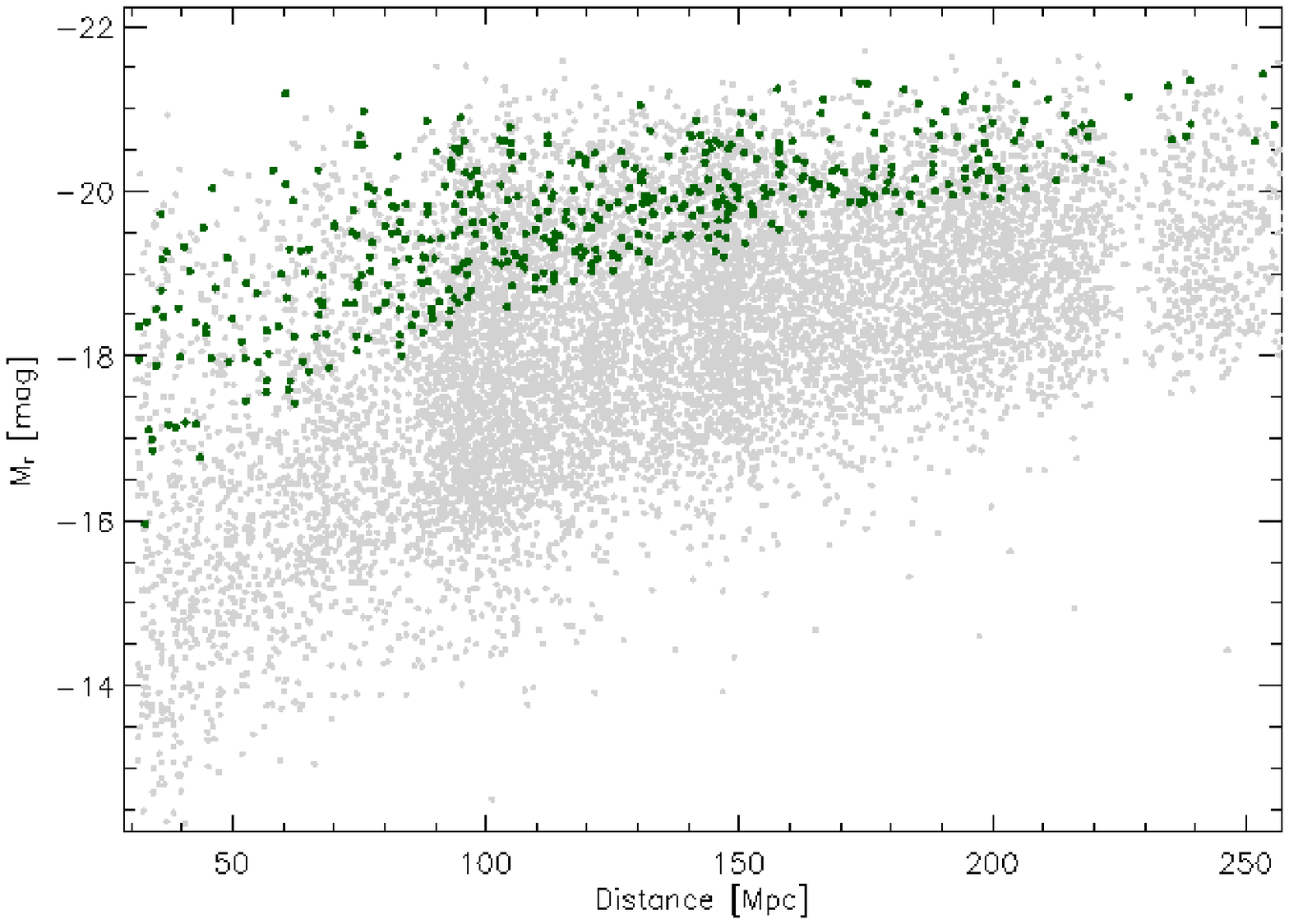} &
\includegraphics[width=0.5\textwidth]{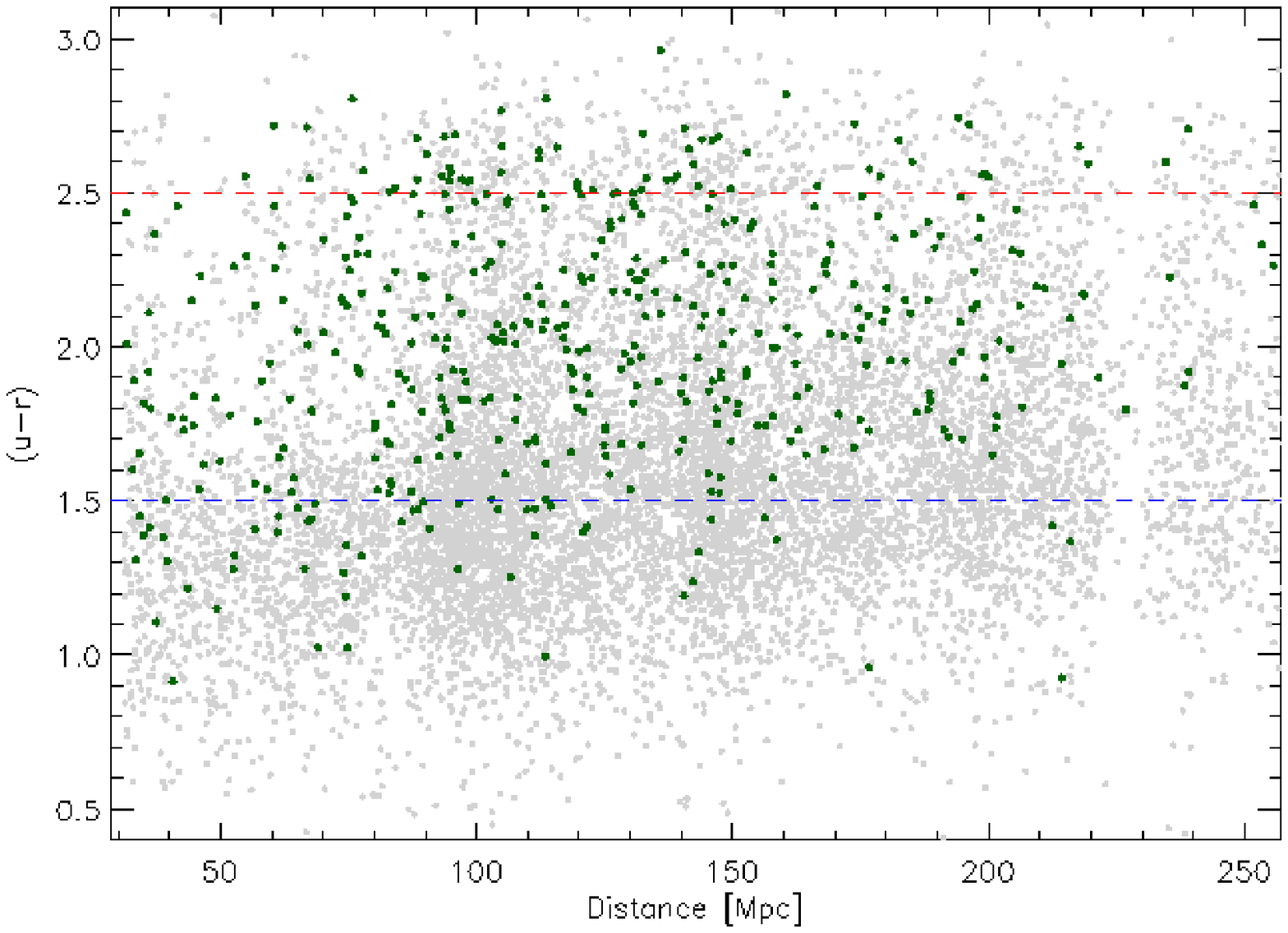}\\

 \includegraphics[width=0.5\textwidth]{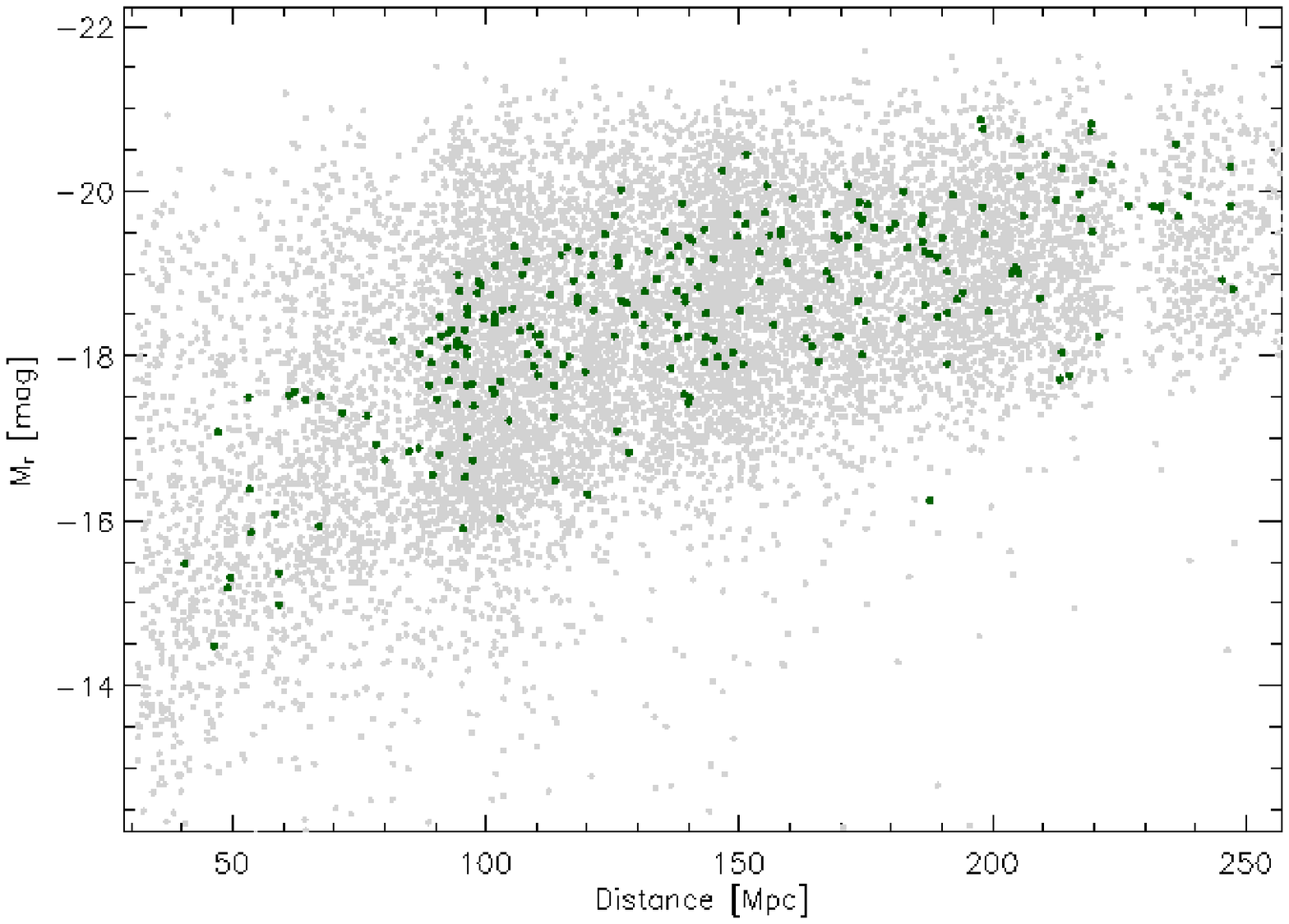} &
 \includegraphics[width=0.5\textwidth]{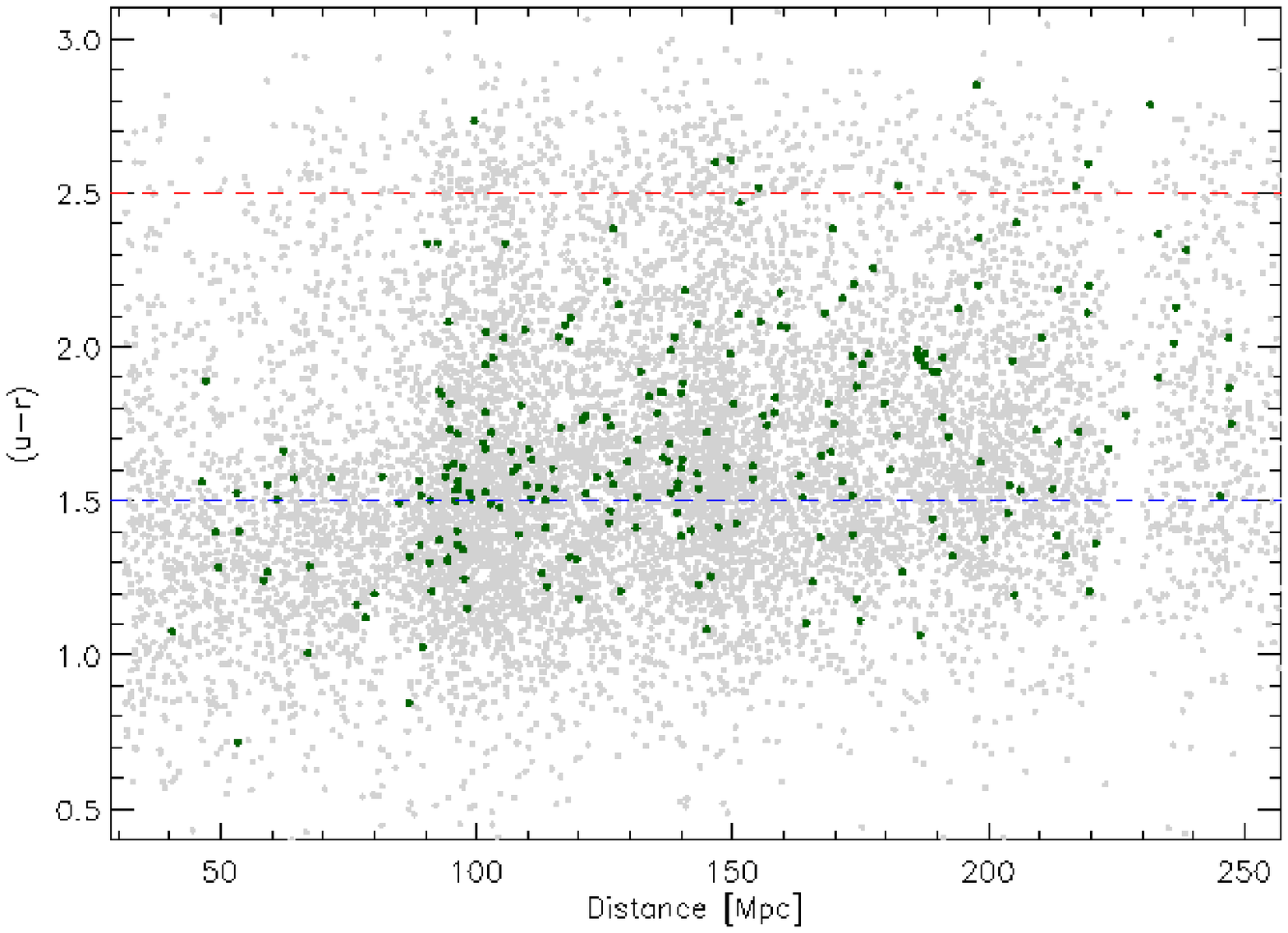}\\

\end{tabular}

\caption{Absolute magnitude $M_r$ (left) and $(u-r)$ color (right)
  vs.\ CMB distance for the three \alf\ control samples: LDE (top),
  IG1 (middle), and IG2 (bottom). The member galaxies are drawn as
  green dots, whereas the middle and bottom plots also depict all the
  \alf\ detections as gray dots. The horizontal dashed lines in the
  $(u-r)$ color plots show the peak values corresponding to the red
  and blue populations (see also Figures~\ref{fig_color-LDE} and
  \ref{fig_color-isolate}).}
\label{fig_dist_alfa_samples}
\end{figure}

These discrepancies, however, do not stem from fundamental differences
between the properties of the objects belonging to the IG1 and IG2
samples, but from the distinct selection criteria adopted to draw up
their member galaxies. As illustrated by the top panels of
Figure~\ref{fig_dist_alfa_samples}, while the luminosity distribution
of the LDE galaxies (as well as that of the full \alf\ catalog) moves
clearly towards brighter values with increasing distance, the
corresponding color distribution, which can be reasonably approximated
by a unimodal function peaking around the characteristic color of the
blue sequence and having a long red tail (see also
Figure~\ref{fig_color-LDE}), shows a modest variation. This rough
distance-invariance is reinforced by the lack of a substantial
red-sequence component for the \hi\ emitters, which start contributing
significantly to the color distribution of the entire galactic
population at somewhat larger redshifts than the blue one. On the
other hand, the isolation selection techniques we used place limits on
the apparent brightness that the candidate galaxies can have. This
inevitably leads to choosing from among the brightest objects that are
within the allowed magnitude range and, consequently, also among the
reddest. Such a bias is stronger in the case of our
spectro-photometric IG1 sample, which excludes objects fainter than
$r=15.27$ mag. As a result, most of the \hi\ sources around the peak
of the blue sequence are missing from the color distribution of this
subset, which only preserves its red tail (middle right panel of
Figure~\ref{fig_dist_alfa_samples}). In contrast, the almost disjoint
constraint $15.5 \leq g_i \leq 19.5$ mag adopted for the definition of
the IG2 sample from the relatively shallow \alf\ data set has led us
to select galaxies that are just slightly intrinsically brighter and,
therefore, slightly redder too, than the average LDE members (bottom
panels of Figure~\ref{fig_dist_alfa_samples}). Note, however, that the
application of the photometric isolation prescription by \citet{All05}
to the much deeper SDSS DR1 catalog led these authors to obtain an
isolated galaxy sample with a light concentration index distribution
that suggested a fifty-fifty morphological composition between early
and late galaxy types, closer to what we observe here for the IG1
subset.

\section{SUMMARY AND CONCLUSIONS}\label{summary}

We have used measurements in the 21-cm emission line from the \alf\
blind survey in a region of the sky also scanned by the SDSS DR7 in
order to study the physical properties of gas-rich galaxies
expected to show little or no evidence of interaction with their
surroundings. The goal has been twofold: 1) to improve the
understanding of the nature of extragalactic \hi\ sources; and 2) to
identify an observational sample of gaseous galaxies as unperturbed as
possible by the environment and therefore suitable to set up standards
for their neutral gas content.

Different parametric estimates of the environment of \hi\ sources have
been explored, in order to establish their suitability in providing a
catalog of gaseous systems whose properties are minimally affected by
external influences. Among the environmental measures tested, the
local number density estimate based on the distance to the 6th-nearest
spectroscopic SDSS neighbor of a galaxy has emerged as the most
appropriate. By applying this estimator to the $11,239$ \hi\
detections identified in the overlapping region of the radio and
optical surveys, we have extracted a subset of 5647 \hi\ sources in
low density environments. We have also defined two different subsets
of isolated galaxies by applying standard isolation criteria based
either on the combination of spectroscopic and photometric information
or solely on photometric data. In spite of the marked differences in
size and membership between the LDE and isolated subsets, their
respective \hi\ content distributions have proven to be all pretty
similar.

In combination with SDSS data, we have studied the distributions of
light concentration indexes, colors, and other proxies of morphology
for the members of these three control samples. All the evidence
analyzed indicate that \hi\ emission is detected essentially in
objects that are structurally similar to galaxies with late-type
optical characteristics. Thus, over $90\%$ of \hi\ sources residing in
a quite environment obey the disk selection rule by
\citeauthor{Mal09}, while some 82 per cent have inverse light
concentration indexes representative of this class. These fractions
could be considered actually a lower limit, as the determination of
the structural parameters of distant objects is quite
imprecise. Likewise, from evidence gathered by studying the color
distributions, we have found that at least between $85\%$ and $90\%$
of the \alf\ detections outside overdense regions are classified as
blue, star-forming galaxies. These fractions are even higher when
extinction-corrected colors are used. In contrast, the structural
properties and colors of \hi\ non-detections are typical of a
population substantially richer in early-type, red objects. The
assignment of morphologies to a subset of \alf\ galaxies in LDEs from
the GZ1 catalog corroborates these conclusions, as only a very small
fraction ($\sim 1$ per cent) of the objects studied have been
classified into the elliptical class with high reliability. These
results suggest that the more reliable automated way to determine
whether a galaxy has genuinely an early- or late-type morphology is by
measuring its \hi\ emission.

The standard view that red galaxies and, consequently, passive objects
of type S0/Sa or earlier, tend to have less neutral hydrogen than
bluer, star-forming late-type systems is therefore confirmed. Our
research also corroborates the prediction, based on the positive tight
correlation between global color and intrinsic luminosity inferred
from optically-selected galaxies, that the fraction of red-/early-type
galaxies in our SDSS-spec LDE sample should increase with increasing
radial velocity of the sources. We observe, however, that while the
color distribution of the objects undetected in \hi\ clearly becomes
redder, that of the gas-rich population does not change significantly
as a function of redshift. This suggests that the color and gaseous
mass of field \hi\ emitters are not very strongly
interrelated. Likewise, we have shown that the close connection
between color and luminosity can have an important impact on isolation
criteria, favoring the selection of the redder/earlier members of a
catalog, the shallower the survey the most notorious the
bias. Awareness of this selection effect, together with the relatively
modest sizes and reduced dynamic range subtended by the properties of
the members of our isolated galaxy samples ---which are a direct
consequence of the severe restrictions imposed by the selection
algorithms on the range of apparent magnitudes---, leads us to
conclude that the LDE approach provides the most suitable and
statistically sounder control sample to set standards of reference for
the \hi\ content of galaxies. In Paper II we will present a detailed
derivation of such standards by applying a non-parametric principal
component analysis to a flux-limited subset of this latter dataset.

\begin{acknowledgements}
\small

This work was supported by the Direcci\'on General de Investigaci\'on
Cient\'{\i}fica y T\'ecnica, under contract AYA2007-60366. M.C.T.\
acknowledges support through a fellowship from the Spanish Ministerio
de Educaci\'on y Ciencia. RG and MPH receive support from NSF grant
AST-0607007 and from a grant from the Brinson Foundation. KLM
acknowledges funding from the Peter and Patricia Gruber Foundation as
the 2008 IAU Fellow, and from the University of Portsmouth and SEPnet
(www.sepnet.ac.uk).

We thank the many members of the \alf\ team who have contributed to
the acquisition and processing of the \alf\ data set over the last six
years.

This research is based mainly on observations collected at Arecibo
Observatory. The Arecibo Observatory is part of the National Astronomy
and Ionosphere Center, which is operated by Cornell University under a
cooperative agreement with the National Science Foundation.

Some of the results quoted in this publication have been made possible
by the participation of more than 160,000 volunteers in the Galaxy Zoo
project. Their contributions are individually acknowledged at
\texttt{http://www.galaxyzoo.org/Volunteers.aspx}.

We have also made use of the HyperLeda database
(\texttt{http://leda.univ-lyon1.fr}) and the NASA/IPAC Extragalactic
Database, which is operated by the Jet Propulsion Laboratory,
California Institute of Technology, under contract from the National
Aeronautics and Space Administration. Likewise, we are grateful to all
the people and institutions that have made possible the NYU
Value-Added Galaxy Catalog
(\texttt{http://sdss.physics.nyu.edu/vagc/}) and the Sloan Digital Sky
Survey (SDSS; \texttt{http://www.sdss.org/}). Funding for the SDSS is provided by the Alfred P.\ Sloan Foundation, the Participating
Institutions, the National Science Foundation, the U.S.\ Department of
Energy, the National Aeronautics and Space Administration, the
Japanese Monbukagakusho, the Max Planck Society, and the Higher
Education Funding Council for England. 

\end{acknowledgements}

\begin{deluxetable}{r|rrr|rr}
\tabletypesize{\scriptsize}
\tablecolumns{6}
\tablewidth{0pt}
\tablecaption{Distributions of ALFALFA and SDSS Galaxies in Radial Velocity Bins\label{tab_alfalfa_detection_frac}}
\tablehead{
\colhead{$\Delta\vhel$}& \multicolumn{1}{|c}{ALFA}& \colhead{ALFA/SDSS-phot}& \colhead{ALFA/SDSS-spec*}& \multicolumn{1}{|c}{SDSS-spec}& \colhead{SDSS-spec/ALFA}\\
\colhead{(km\ s$^{-1}$)} & \multicolumn{1}{|c}{$N_{\rm gal}$} & \colhead{$N_{\rm gal}$ ($\%$)} & \colhead{$N_{\rm gal}$ ($\%$)} & \multicolumn{1}{|c}{\ $N_{\rm gal}$} & \colhead{$N_{\rm gal}$ ($\%$)} 
}
\startdata

$<500$\ \ \           &   71  &   65 (91\%)\ \ \ \  &    50 (77\%)\ \ \ \ &    37  &   18 (49\%)\ \ \\
500--4000\ \ \        & 1523  & 1417 (93\%)\ \ \ \  &   1024 (72\%)\ \ \ \  &  1702  &  910 (53\%)\ \ \\
4000--7500\ \ \       & 3104  & 3066 (99\%)\ \ \ \  &  2529 (82\%)\ \ \ \  &  5102  & 2420 (47\%)\ \ \\
7500--11,000\ \ \     & 3385  & 3353 (99\%)\ \ \ \  &  2947 (88\%)\ \ \ \  &  9196  & 2853 (31\%)\ \ \\
11,000--14,500\ \ \   & 2280  & 2261 (99\%)\ \ \ \  &  2025 (89\%)\ \ \ \  &  9341  & 1977 (21\%)\ \ \\\ \ \ \ 
14,500--18,000\ \ \   &  876  &  871 (99\%)\ \ \ \  &   807 (93\%)\ \ \ \  &  10187  &  787 (\,\ 8\%)\ \ \\
\hline
Totals\ \ \           & 11,239  & 11,033 (98\%)\ \ \ \  &  9382 (85\%)\ \ \ \  & 35,565 & 8965 (25\%)\ \ \\

\enddata

\tablecomments{Numbers in different velocity bins of ALFALFA galaxies
  (ALFA), \smallhi\ detections with a photometric counterpart in SDSS
  (ALFA/SDSS-phot), and \smallhi\ detections with a spectroscopic
  counterpart in SDSS (ALFA/SDSS-spec*), regardless of the
  morphological classification of their SDSS spectrum, in the region
  where both catalogs overlap. SDSS-spec designates the sample of SDSS
  sources with both photometric and spectral morphology classified as
  of galaxy type and having $r < 17.77$ mag, while SDSS-spec/ALFA
  identifies the subset of SDSS-spec galaxies with an ALFALFA
  counterpart. Fractions are given in parentheses. }

\end{deluxetable}

\begin{deluxetable}{l|r|l}
\tabletypesize{\scriptsize}
\tablecolumns{3}
\tablewidth{0pt}
\tablecaption{\hi\ Control Samples\label{tab_samples}}
\tablehead{\colhead{Acronym}& \multicolumn{1}{|l}{$N_{\rm gal}$} &  \multicolumn{1}{|c}{Selection criterion} }
\startdata
LDE\ \ &5647& Low local galaxy density environment ($\rho_6\leq 0.5\,h^3$ galaxies~Mpc$^{-3}$)\\
IG1\ \ & 461& Isolation inferred from spectro-photometric information (Equation~\ref{IG1})\\
IG2\ \ & 236& Isolation inferred from photometric information (Equation~\ref{IG2})\\
\enddata

\end{deluxetable}

\begin{deluxetable}{r|rr|rr}
\tabletypesize{\scriptsize}
\tablecolumns{5}
\tablewidth{0pt}
\tablecaption{ALFALFA Detections and SDSS Galaxies in LDEs Obeying \citeauthor{Mal09}'s Criterion\label{tab_maller}}
\tablehead{
\colhead{$\Delta\vhel$}&\multicolumn{2}{|c|}{SDSS-spec} & \multicolumn{2}{c}{SDSS-spec/ALFA}\\   
\colhead{(km\ s$^{-1}$)} &  \multicolumn{1}{|c}{$N_{\rm gal}$} & \colhead{$N_{\rm disk}$ ($\%$)}&  \multicolumn{1}{|c}{$N_{\rm gal}$} & \colhead{$N_{\rm disk}$ ($\%$)}
}
\startdata
2000--7000\ \ &2063 &1893 ($91.7\%$)& 1330 &1257 ($94.5\%$)\\
7000--11,000\ \ &3912&3403 ($87.0\%$)& 1529  &1411 ($92.3\%$)\\
11,000--18,000\ \ &7889&6472 ($82.0\%$)& 1555&1414 ($90.9\%$)\\
\hline
Totals \ \  & 13864& 11768 ($84.9\%$)& 4414 &4082 ($92.5\%$)\\
\enddata

\tablecomments{Numbers in different velocity bins of SDSS-spec and
  SDSS-spec/ALFA galaxies (the definition of these subsets can be
  found in the caption of Table~\ref{tab_alfalfa_detection_frac}) in
  low density environments satisfying \citeauthor{Mal09}'s
  disk-identification criterion ($N_{\rm disk}$). We only consider
  galaxies with ($r$-band) S\'ersic index $0.5<n<5.9$ and axis ratio
  $0.15<b/a<1$. Fractions are given in parentheses.}

\end{deluxetable}

\clearpage

\end{document}